\documentclass[a4paper,10pt]{article}

\usepackage{amssymb,amsmath,xcolor,color,cite,graphicx,soul}

\title{Super-Klein tunneling of Dirac fermions through electrostatic gratings  in graphene }
\author{Alonso Contreras-Astorga$^1$, Francisco Correa$^{2}$, V\'i{}t Jakubsk\'y$^3$\\\vspace{1mm}\\  \vspace{1.8mm}
	$^1$\textit{CONACYT-Departamento de F\'isica, Cinvestav, A.P. 14-740, 07000 Ciudad de M\'exico, Mexico}\\
	$^2$\textit{ Instituto de Ciencias F\'isicas y Matem\'aticas, }\\ \vspace{1.8mm}
\textit{Universidad Austral de Chile, Casilla 567, Valdivia, Chile}\\ 
	$^3$\textit{The Czech Academy of Science, Nuclear Physics Institute, \v Re\v z/Prague, Czech
		Republic}\\
	\\\sl{\small{E-mails: alonso.contreras@conacyt.mx, francisco.correa@uach.cl,  
			jakubsky@ujf.cas.cz}} }

\begin{document}
	\maketitle
\begin{abstract}
	\begin{itemize}
	We use the Wick-rotated time-dependent supersymmetry to construct models of two-dimensional Dirac fermions in presence of an electrostatic grating. We show that there appears omnidirectional perfect transmission through the grating at specific energy. Additionally to being transparent for incoming fermions, the grating hosts strongly localized states. 
		\end{itemize}
 \end{abstract}

\section{Introduction}
Super-Klein tunneling (also called all-angle or omnidirectional Klein tunneling) is a phenomenon where relativistic particles can go through an electrostatic barrier at any angle without being reflected. It was theoretically predicted for quasi-particles of spin-$1$ in Dice lattices \cite{urban}, in line-centered square lattices formed by ultracold atoms \cite{shen}, in optics \cite{louie}, or in the 
systems with spin-orbit interaction \cite{betancur1}. It was also analyzed for the spin-$1/2$  Dirac fermions in graphene \cite{betancur1/2} with an inhomogeneous Fermi velocity  as well as for the spin-less particles described by the Klein-Gordon equation \cite{kim}.  
Omnidirectional Klein tunneling occurs exclusively for a specific energy. When the energy of the incoming particle is mismatched with this critical value, transmission gets strongly angle-dependent and the perfect tunneling takes place for some sharp values of incidence angle only \cite{kng}. For instance, the total transmission occurs for massless fermions bouncing in normal direction on the barrier that has translational symmetry. In this case, the absence of backscattering can be explained by the relation of the setting with the the free-particle model \cite{disguise}.

The all-angle Klein-tunneling was studied in models with piece-wise constant potentials, typically rectangular barriers, that possessed translation invariance in one direction  \cite{urban,shen,louie,betancur1,betancur1/2}. The explicit solutions and scattering amplitudes (transmission coefficient in particular) were found for arbitrary energies. Then it was shown that for a specific energy 
there appears total transmission that is independent on the incidence angle.

In general, the Klein tunneling makes it difficult to confine Dirac fermions using electrostatic fields. There is much effort in the literature to show that despite Klein tunneling, the confinement of Dirac fermions by electrostatic fields is actually possible. Guided modes in waveguides were revealed in \cite{hartmann,hartmann2,vpn1} for systems with translation symmetry, see also the rigorous results in \cite{elton}. Quasi-bound states in quantum dots were considered in \cite{matulis} whereas square integrable bound states were found in \cite{bardarson,vpn2,portnoizero} for systems with rotational symmetry. In these works, both analytic and numerical calculations were employed to find the localized states. The square integrable wave functions were usually zero-energy modes. 

Supersymmetric quantum mechanics was used in the seek for  localized zero-energy solutions of the two-dimen\-si\-onal Dirac equation with electrostatic potential, \cite{ioffe1,ioffe2,axel1,axel2,royho,ghoshroy}. In \cite{axel1}, \cite{royho}, \cite{ghoshroy} translation symmetry allowed to deal with effectively one-dimensional Dirac operator. Its square was identified with a known exactly solvable nonrelativistic model and zero-modes were found analytically. In \cite{axel2}, supersymmetric transformations were studied for two-dimensional Dirac operators with matrix potential that possessed translation symmetry. Existence of zero modes for Dirac Hamiltonians in presence of the electrostatic potentials on a bounded domain was discussed in \cite{ioffe1}. Generalized intertwining relations for Dirac operator with electrostatic potential were used for construction of truly two-dimensional settings that possessed zero-modes, see \cite{ioffe2}. 

In the this article, we present models where the super-Klein tunneling phenomenon occurs in presence of  smooth electrostatic potential that lacks translation symmetry and resembles a grating formed by a periodic chain of localized scatterers. We provide exact solutions of the corresponding stationary Dirac equation for a fixed value of energy where the super-Klein tunneling occurs.  We discuss scattering properties and show that there are also states that are strongly localized by the electrostatic field. 

The article is organized as follows. In the next section, we make a brief review of the time-dependent supersymmetric transformation for $1+1$ dimensional Dirac equation and present two explicit models. In section three, we discuss Wick rotation of $1+1$ dimensional Dirac equation that results in $0+2$ dimensional stationary Dirac equation. 
In the section four, we analyze the explicit models, their scattering properties and existence of localized states at a fixed, non-zero energy. 
We show that the reflection is missing in the considered regime and the omnidirectional Klein tunneling appears naturally. The last section is left for discussion.

\section{Time-dependent SUSY transformations}
Supersymmetric quantum mechanics provides a framework for construction of new exactly solvable models from known ones \cite{khare}. It relies on the supersymmetric (susy) transformation $L$ that intertwines the operator $H_0$, be it a Schr\"odinger or a Dirac operator of a known model, with an operator $H_1$ that represents the new quantum system,
\begin{equation}\label{genintertwining}
LH_0=H_1L,\quad L^{\dagger}H_1=H_0L^{\dagger}.
\end{equation}  
The second relation can be obtained by conjugation of the first one provided that $H_0$ and $H_1$ are Hermitian.
The intertwining operator is known for a long time in the analysis of differential equations as Darboux transformation, see e.g. \cite{matveev} and references therein. The intertwining relations for the stationary one-dimensional Dirac equation were discussed in \cite{SamsonovStac}, for the non-stationary one in \cite{Samsonov}, see also \cite{NDSchulze} where the higher dimensional case was considered. 

The operator $L$ is usually defined as a differential operator that annihilates a selected set of eigenvectors of $H_0$. The choice of these ``seed" vectors also determines the explicit form of $H_1$. It has major influence on Hermiticity of $H_1$ and regularity of its potential term.  The intertwining relations are quite powerful; when $\psi$ satisfies $H_0\psi=0$, then $L\psi$ is a solution of $H_1L\psi=0$. Since the relations (\ref{genintertwining}) are formal, one has to guarantee that the physical states of $H_0$ are mapped  properly into the physical states of $H_1$ by the operator $L$.

In this section, we briefly review the time-dependent susy transformation for the $1+1$ dimensional Dirac equation, see \cite{Samsonov}  for more details, and present two explicit models that will be further elaborated in the next section. 
Let us have the initial equation in the following form,
\begin{equation}\label{td}
H_0\psi=(i\partial_t-i\sigma_2\partial_z-\mathbf{V}_0(z,t))\psi=0,
\end{equation}
where the matrix potential $\mathbf{V}_0(z,t)$ is supposed to be Hermitian. The susy transformation is based on the choice of two vectors $u_1(z,t),$ $u_2(z,t)$ that are solutions of (\ref{td}). We can use them to compose a matrix  $\mathbf{u}=(u_1,u_2)$ that, by definition, satisfies 
\begin{equation}
H_0\mathbf{u}=0.
\end{equation}
Once the matrix $\mathbf{u}$ is fixed, it is possible to define intertwining operators $L$, $L^{\dagger}$ and the new Dirac operator $H_1$ in the following manner,
\begin{eqnarray}\label{L}
L&=&\partial_z-\mathbf{u}_z\mathbf{u}^{-1},\quad L^\dagger=-\partial_z-(\mathbf{u}^\dagger)^{-1}\mathbf{u}^\dagger_z,\\ \label{acheuno}
H_1&=&H_0-i[\sigma_2,\mathbf{u}_z\mathbf{u}^{-1}]=i\partial_t-i\sigma_2\partial_z-\mathbf{V}_0(z,t)-\mathcal{V}^{(1)}_1(z,t)\sigma_1-\mathcal{V}^{(1)}_3(z,t)\sigma_3.
\end{eqnarray}
It can be checked by direct calculation that the operators $L$, $L^{\dagger}$ and $H_1$ satisfy (\ref{genintertwining}), see \cite{Samsonov}. The choice of the seed solutions $u_1(z,t)$ and $u_2(z,t)$ determines the properties of the potential term in $H_1$.
It is usually required fo fix the seed solutions such that $\mathcal{V}^{(1)}_1(z,t)$ and $\mathcal{V}^{(1)}_3(z,t)$ are real and regular. This requirement should guarantee that physical eigenstates of $H_0$ are mapped by $L$ into the physical eigenstates of $H_1$.  

By definition, the matrix $\mathbf{u}$ gets annihilated by the intertwining operator $L$, $L\mathbf{u}=0$. The conjugate operator $L^{\dagger}$ annihilates a matrix $\mathbf{v}$ that satisfies
\begin{equation}\label{v}
\mathbf{v}=(\mathbf{u}^{\dagger})^{-1},\quad L^{\dagger}\mathbf{v}=0,\quad H_1\mathbf{v}=0.
\end{equation}
Therefore, the columns $v_1$ and $v_2$ of $\mathbf{v}=(v_1,v_2)$ are solutions of the equation $H_1v_a=0$, $a=1,2$. 

Let us illustrate the framework now on two explicit examples.

\subsection{Model A}
We start by fixing the free-particle model as the initial system. The dynamical free equation can be written as
\begin{equation}\label{free}
H_0\psi=(i\partial_t-i\sigma_2\partial_z-m\sigma_3)\psi=0,
\end{equation}
where $m$ is a real parameter. It is straightforward to find its explicit solutions. 
They can be written as
\begin{equation}\label{basis}
\psi_\pm(z,t)= e^{\pm k z}\left(\mp \frac{m \cosh ( \omega t)+i \omega  \sinh ( \omega t)}{k}, \cosh ( \omega t)\right)^T \, ,
\end{equation}
where $k=\sqrt{\omega^2+m^2}$. In order to proceed with time-dependent susy mechanism, we define the matrix $\mathbf{u}$ in this model as $\mathbf{u}=(\psi_1,\sigma_1\psi_1^*)$ where
\renewcommand\arraystretch{1}
\begin{eqnarray}\label{modelApsi1}
\psi_1(z,t)
&=&\frac{1}{2}\left(\psi_+(z,t)+\psi_-(z,t) \right)\\&=&\left(-(m\cosh (\omega t)+i\omega \sinh (\omega t))\frac{\sinh (kz) }{k}, \cosh(\omega t)\cosh (kz)\right)^T\, .
\end{eqnarray}
\renewcommand\arraystretch{1}
By virtue of Eq. (\ref{L}), the time-dependent intertwining operators $L(z,t)$ and $L^\dagger(z,t)$ read as
\begin{eqnarray}
L&=&\partial_z-\frac{1}{2D_1}\left(\frac{\omega^2\sinh(2kz)}{k}\sigma_0-2m\cosh^2 (\omega t)\, \sigma_1+  \omega \sinh  (2\omega t) \,\sigma_2\right),\\
L^\dagger&=&-\partial_z-\frac{1}{2D_1}\left(\frac{\omega^2\sinh(2kz)}{k}\sigma_0-2m\cosh^2 (\omega t) \,\sigma_1+\omega 
\sinh (2\omega t) \, \sigma_2\right).
\end{eqnarray} 
Here, we abbreviated $D_1(z,t)=(m^2+k^2\cosh(2\omega t)+\omega^2\cosh (2kz))/2 k^2$. We can construct the new Dirac operator $H_1$ from the relation (\ref{acheuno}),
\begin{equation}\label{h1}
H_1=i\partial_t-i\sigma_2\partial_z+\left(-m+\frac{4mk^2\cosh^2 (\omega t)}{m^2+k^2\cosh (2\omega t)+\omega^2\cosh (2k z)}\right)\sigma_3.
\end{equation}
It is intertwined with $H_0$ by $L$ via the Eq. (\ref{genintertwining}). The potential term represents a fluctuation of the constant mass,  exponentially localized both in time and space. In contrast with the original, free-particle system (\ref{free}), the new system possesses two solutions which are localized in space-time.  The vectors $v_1$ and $v_2$, see (\ref{v}), satisfy $H_1v_1=H_1v_2=0$ and have the following explicit form,
\begin{eqnarray}\label{G}
v_1(z,t)&=&\frac{1}{D_1}\left([m\cosh (\omega t)+i\omega \sinh (\omega t)]\sinh kz,k\cosh(\omega t)\cosh (zk)\right)^T,\nonumber\\
v_2(z,t)&=&\sigma_1 v_1(z,t)^*.
\end{eqnarray}
\subsection{Model B}

In the second example, we identify $H_0$ with the free-particle operator (\ref{free}) again. We fix the following linear combination of the solutions of (\ref{free}),
\begin{equation}\label{modelBpsi2}
\psi_2=\frac{i}{2k}\sigma_1 \psi_+^*\left(x,t+\frac{i \pi}{2\omega}\right)+\frac{i k s}{2\omega^2}\sigma_1 \psi_-^*\left(x,t+\frac{i \pi}{2\omega}\right)-\frac{k^2}{\omega^2}(c_2+i c_1) \psi_-\left(x,t+\frac{i \pi}{2\omega}\right)\, ,
\end{equation}
where  we used the basis states (\ref{basis}) and $s$, $c_1$ and $c_2$ are constants. As in the previous case, we build the matrix $\mathbf{u}=(\psi_2,\sigma_1\psi_2^*)$ that  satisfies by definition  $H_0\mathbf{u}=0$.  
The corresponding intertwining operators are
\begin{eqnarray}\label{LmodelB}
L = \partial_z - \frac{1}{D_2}\left( \ell_0 \sigma_0 + \ell_1 \sigma_1 + \ell_2 \sigma_2 \right),    \quad
L^\dag =  -\partial_z - \frac{1}{D_2}\left( \ell_0 \sigma_0 + \ell_1 \sigma_1 + \ell_2 \sigma_2 \right),
\end{eqnarray}
where
\begin{align}
&\ell_0(z)  = \omega^4  k e^{2 k z}{-}k^5\left[s^2{-}4(c_1^2{+}c_2^2)k^2  \right] e^{-2kz}, \\
&\ell_1(t) = 2 k^4\left[ ms +2 c_1 k^2 -  m(2 c_1 m + s) \cosh(2\omega t) + 2 c_2 m \omega \sinh(2 \omega t)  \right], \\
&\ell_2(t)  = 2 \omega k^4 \left[(2c_1 m+s) \sinh (2 \omega t )-2 c_2 \omega  \cosh (2  \omega t )\right], \\
&D_2(z,t)=\omega^4 e^{2 k z}{+}k^4\left[s^2{-}4(c_1^2{+}c_2^2)k^2  \right] e^{-2kz}{+}2k^4 \left[ (2 c_1 m {+}s) \cosh(2\omega t){-}2 c_2 \omega \sinh(2\omega t)   \right]{-}2k^2m(m s{+}2 c_1 k^2).
\end{align}

The new Dirac operator $H_1$ follows from Eq. (\ref{acheuno}) and reads
\begin{equation}\label{h2}
H_1=\left(i\partial_t-i\sigma_2\partial_z-\left(m + W(z,t) \right)  \sigma_3 \right).
\end{equation}
The new potential term $W(z,t)$ is explicitly
\begin{equation}\label{W2}
W(z,t)= \frac{2 k^4\left[ 2ms +4 c_1 k^2 - 2 m(2 c_1 m + s) \cosh(2\omega t) + 4 c_2 m \omega \sinh(2 \omega t)  \right] }{D_2(z,t)}. 
\end{equation}
We can see that $W(z,t)$ is real function so that $H_1$ is Hermitian provided that $s$, $c_1$ and $c_2$ are real.
Let us omit here the discussion on regularity of $W(z,t)$  since it is not relevant in this section. It is worth noticing that (\ref{W2}) for $c_1=0$ and $c_2=0$ was discussed in \cite{Samsonov}. In that work, it was obtained via the so called confluent susy transformation that is a specific case of the second-order susy transformation, see also \cite{Mielnik}, \cite{Fernandez04} for more details. 

The eigenstates $v_{1,2}$ of $H_1$ can be found from (\ref{v}). They are 
\begin{eqnarray}\label{G2}
v_1=\frac{2 k^2}{D_2} \left( 
\begin{array}{c}
k^3\left[s \sinh(\omega t)+2(c_1+i c_2)M(t)\right]e^{-kz}-\omega^2 k \sinh(\omega t) e^{kz} \\
k^2\left[i s \omega \cosh(\omega t)-(ms +2 (c_1+ic_2)k^2  ) \sinh(\omega t)  \right] e^{-kz}-\omega^2 M^*(t) e^{kz}  \\
\end{array} 
\right)
\end{eqnarray}
and $v_2=\sigma_1 v_1^*$, where $M(t)=m \sinh(\omega t)+i \omega \cosh(\omega t)$.

\section{Wick rotation}
The Wick rotation was used originally in \cite{Wick} where it was employed to get solutions of Bethe-Salpeter equation in Minkowski space from those defined in the Euclidean space. Let us go in the opposite direction and define a system living in two-dimensional Euclidean space from a related  1+1 dimensional relativistic model and get a relevant information on its physical properties.

We make the following change of the coordinates in the Dirac equation (\ref{td}),
\begin{equation}\label{x->kappa}
z=ix,\quad \partial_z=-i\partial_x,\quad t=y.
\end{equation}
It turns the indefinite Minkowski metric into the euclidean one and transforms the equation (\ref{td}) into   
\begin{equation}\label{dy}
H_0(ix,y)\psi(ix,y)=(i\partial_y-\sigma_2\partial_x-V_0(ix,y))\psi(ix,y)=0.
\end{equation}
It does not have the form of a stationary equation of a two-dimensional system yet. Let us fix it by
multiplying (\ref{dy}) by $\sigma_3$ from the left and making an additional gauge transform $\mathbf{U}=e^{i\frac{\pi}{4}\sigma_1}$. This way, we get the following two-dimensional stationary equation for zero energy in terms of the spatial coordinates $x$ and $y$,
\begin{equation}\label{2d2a}
\widetilde{H}_0(x,y)\widetilde{\psi}(x,y)=-\mathbf{U}\sigma_3H_0(ix,y)\mathbf{U}^{-1}\widetilde{\psi}(x,y)=(-i\sigma_1\partial_x-i\sigma_2\partial_y+\widetilde{\mathbf{V}}_0(x,y))\mathbf{U}\psi(ix,y)=0
\end{equation}
with the potential term 
\begin{eqnarray}\label{v3}
\widetilde{\mathbf{V}}_0(x,y)&=&\mathbf{U}\sigma_3\mathbf{V}_0(ix,y)\mathbf{U}^{-1}.
\end{eqnarray}
The solutions $\psi(z,t)$ of (\ref{td}) transform into the solutions $\widetilde{\psi}(x,y)$ of (\ref{2d2a}),
\begin{equation}\label{tildepsi}
\widetilde{\psi}(x,y)=\mathbf{U}\psi(ix,y).
\end{equation}

A few comments are in order. The operations (\ref{x->kappa})-(\ref{2d2a}) can render the Dirac operator $\widetilde{H}_0$ non-Hermitian in general. As we shall see in the next section, this problem can be successfully addressed in specific examples where it is possible to keep $\widetilde{H}_0$ Hermitian. The transformation also makes profound changes into the character of the potential term. With $\mathbf{V}_0(z,t)=\sum_{a=0}^3\mathcal{V}_a(z,t)\sigma_a$, we get 
\begin{equation}\label{tildeV}
\widetilde{\mathbf{V}}_0(x,y)=\mathbf{U}\sigma_3\mathbf{V}_0(ix,y)\mathbf{U}^{-1}=\sigma_0\mathcal{V}_3(ix,y)-i\sigma_1 \mathcal{V}_2(ix,y)+\sigma_2 \mathcal{V}_0(ix,y)-i\sigma_3 \mathcal{V}_1(ix,y).
\end{equation}
In particular, the mass term $\mathcal{V}_3(z,t)\sigma_3$ of $H_0(z,t)$ turns into the electrostatic potential $\mathcal{V}_3(ix,t)$ of $\widetilde{H}_0(x,y)$. 
The price paid for the Wick rotation leading to (\ref{2d2a}) is the fact that (\ref{2d2a}) describes the system at a single, fixed energy $E=0$. As much as we find this fact restrictive, it can still provide highly non-trivial and valuable physical information on the system as we shall see in the next section.

Let us see how the Wick rotation modifies the intertwining relations and the susy transformation.
Changing the coordinates and multiplying (\ref{genintertwining}) from the left by $\mathbf{U}\sigma_3$ and from the right by $\mathbf{U}^{-1}$, the intertwining relation can be written in the following form,
\begin{align}\label{wrintertwining}
\widetilde{L}^\star (x,y)\widetilde{H}_0(x,y)&=\widetilde{H}_1(x,y) \widetilde{L}(x,y)
\end{align}
where we denoted
\begin{equation}\label{new}
\widetilde{L}=\mathbf{U}L(ix,y)\mathbf{U}^{-1},\quad 
\widetilde{L}^\star(x,y)= \mathbf{U}\sigma_3L(ix,y)\sigma_3\mathbf{U}^{-1}.
\end{equation}
Hence, the intertwining relation gets substantially altered as there appears $\widetilde{L}^\star(x,y)$ on the left and while there is $\widetilde{L}(x,y)$ on the right side and they do not coincide in general. The equation (\ref{wrintertwining}) is a special case of the construction provided recently in \cite{ioffe2} where generalized interwining relations were discussed for zero modes of a two-dimensional Dirac operator.
As we deal with the zero modes of $\widetilde{H}_0$ and $\widetilde{H}_1$  only, we can reduce the intertwining relations into the following relevant formulas (we suppose that $\widetilde{H}_1$ is Hermitian),
\begin{equation}\label{tildeL}
\widetilde{H}_0\Psi_0=0\Rightarrow \widetilde{H}_1\widetilde{L}\Psi_0=0,\quad \widetilde{H}_1\Psi_1=0\Rightarrow \widetilde{H}_0\widetilde{L}^\dagger\Psi_1=0,
\end{equation}
The Dirac operator (\ref{h2}) constructed via the susy transformation and Wick-rotated afterwards reads as
\begin{eqnarray}\widetilde{H}_1(x,y)&=&-\mathbf{U}\sigma_3H_1(ix,y)\mathbf{U}^{-1}\\
&=&-i\sigma_1\partial_x-i\sigma_2\partial_y+\widetilde{\mathbf{V}}_0(ix,y) + \sigma_0 \mathcal{V}^{(1)}_3(ix,y) -i\sigma_3 \mathcal{V}^{(1)}_1(ix,y) .\end{eqnarray}
The last formula reveals that the susy transformation in combination with the Wick rotation adds an electrostatic interaction and a mass term into the new system.

\section{Super-Klein tunneling on the electrostatic grating
}
Let us return now to the $1+1$ dimensional models $A$ and $B$ and apply on them the Wick rotation. We will see that the resulting $0+2$ dimensional systems possess remarkable physical properties that can be investigated with the use of the intertwining operators.
\subsection{Wick-rotated model A}
\begin{figure}\begin{center}
		\includegraphics[scale=.32]{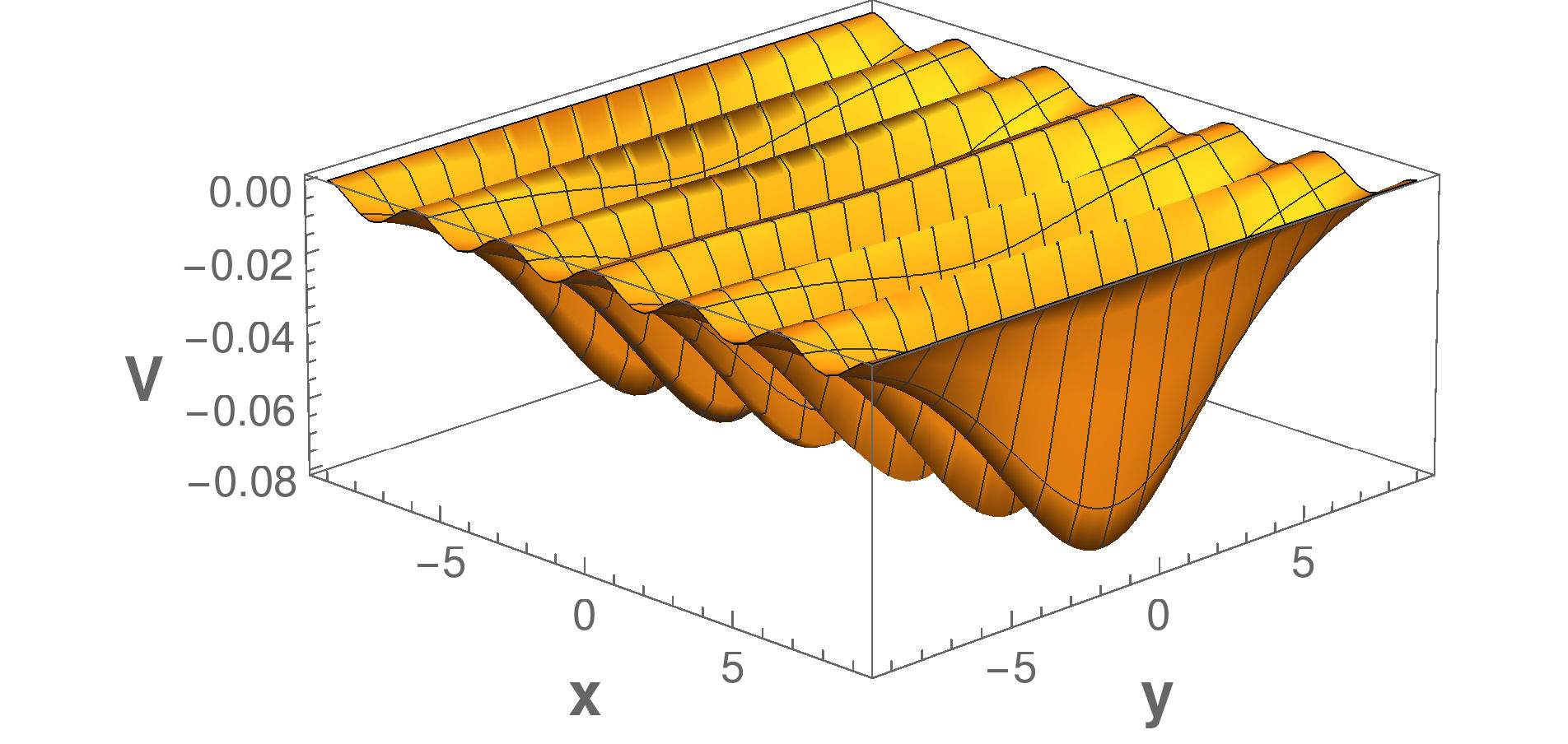}
		\includegraphics[scale=0.32]{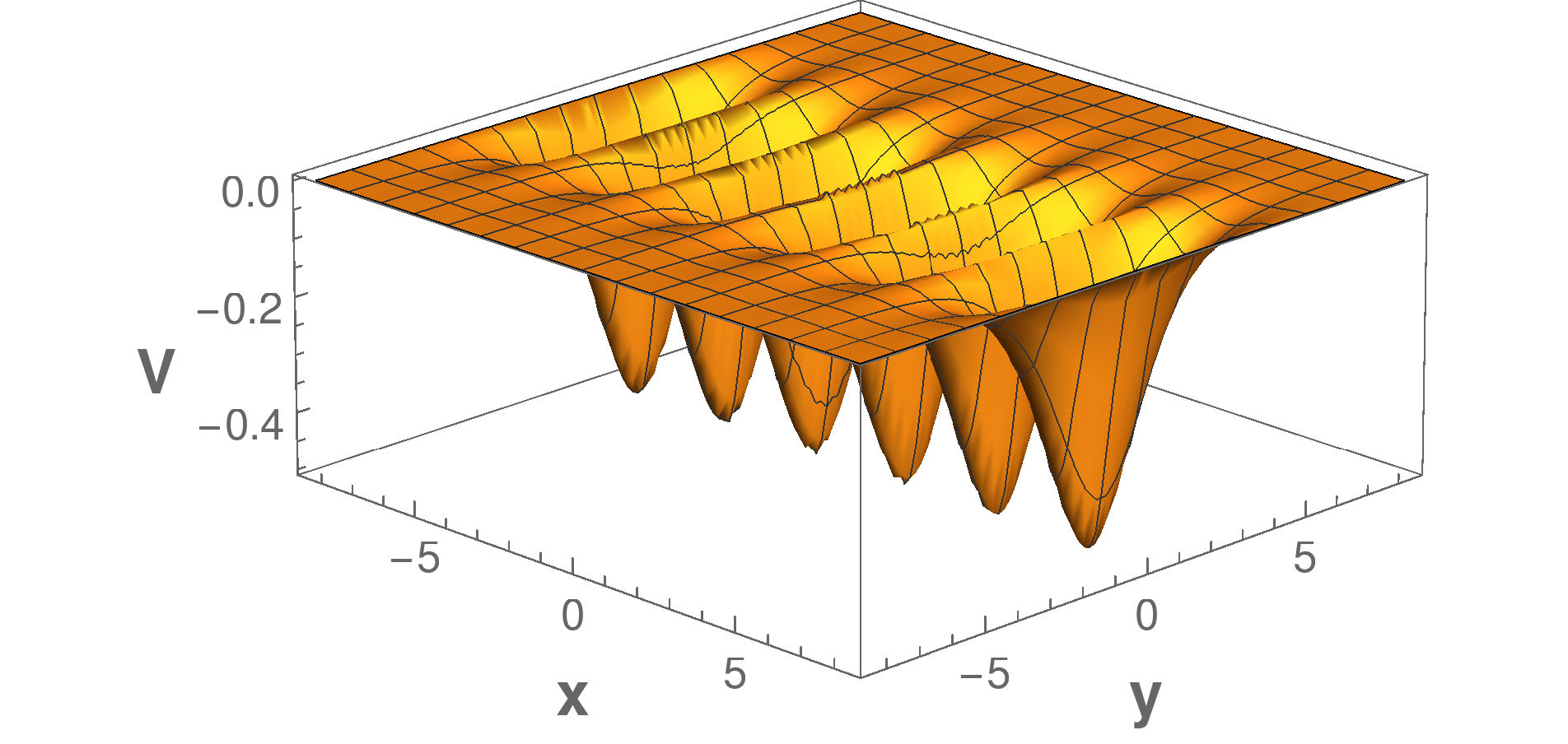}
		\includegraphics[scale=.32]{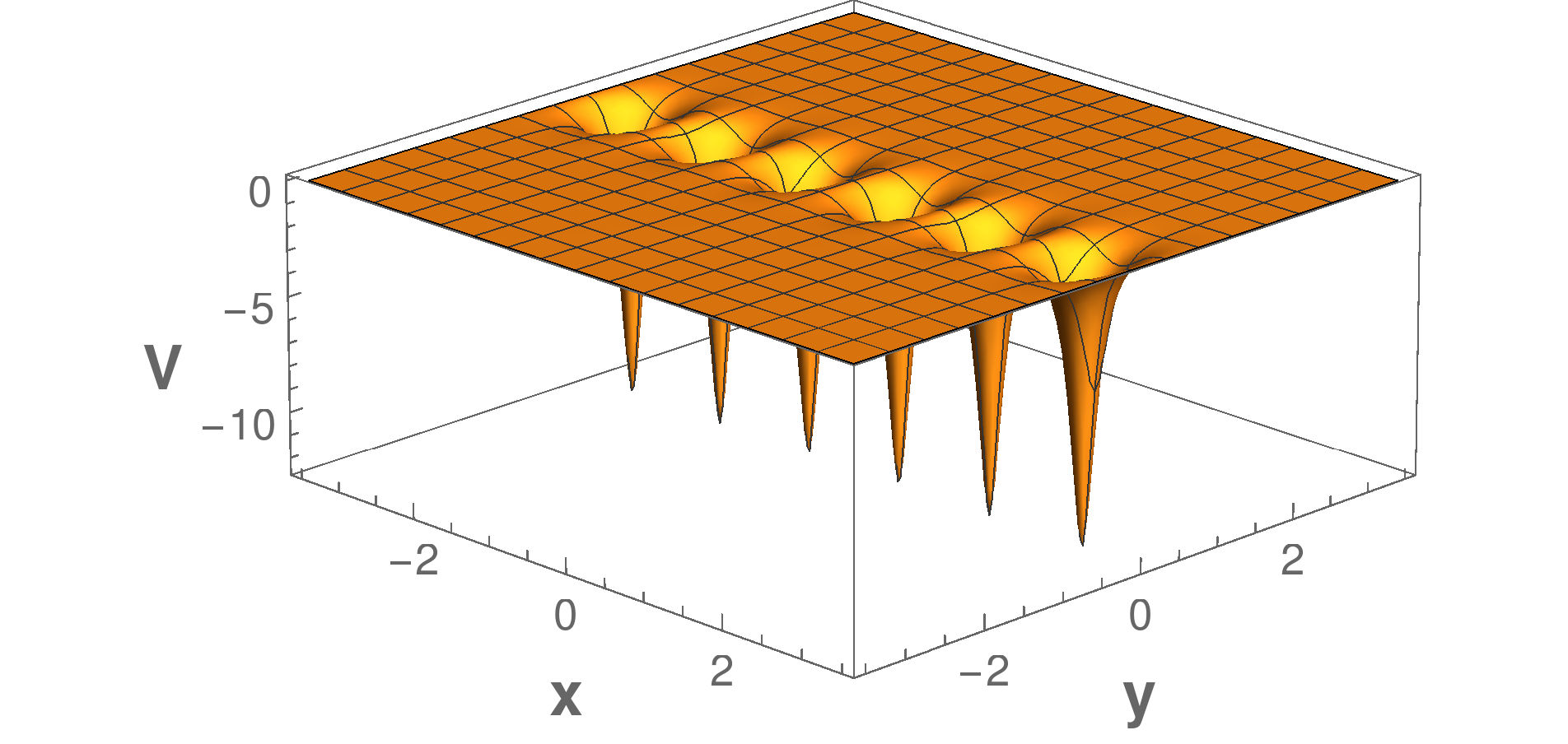}\\
		$ $ $ $\\
		\includegraphics[scale=0.32]{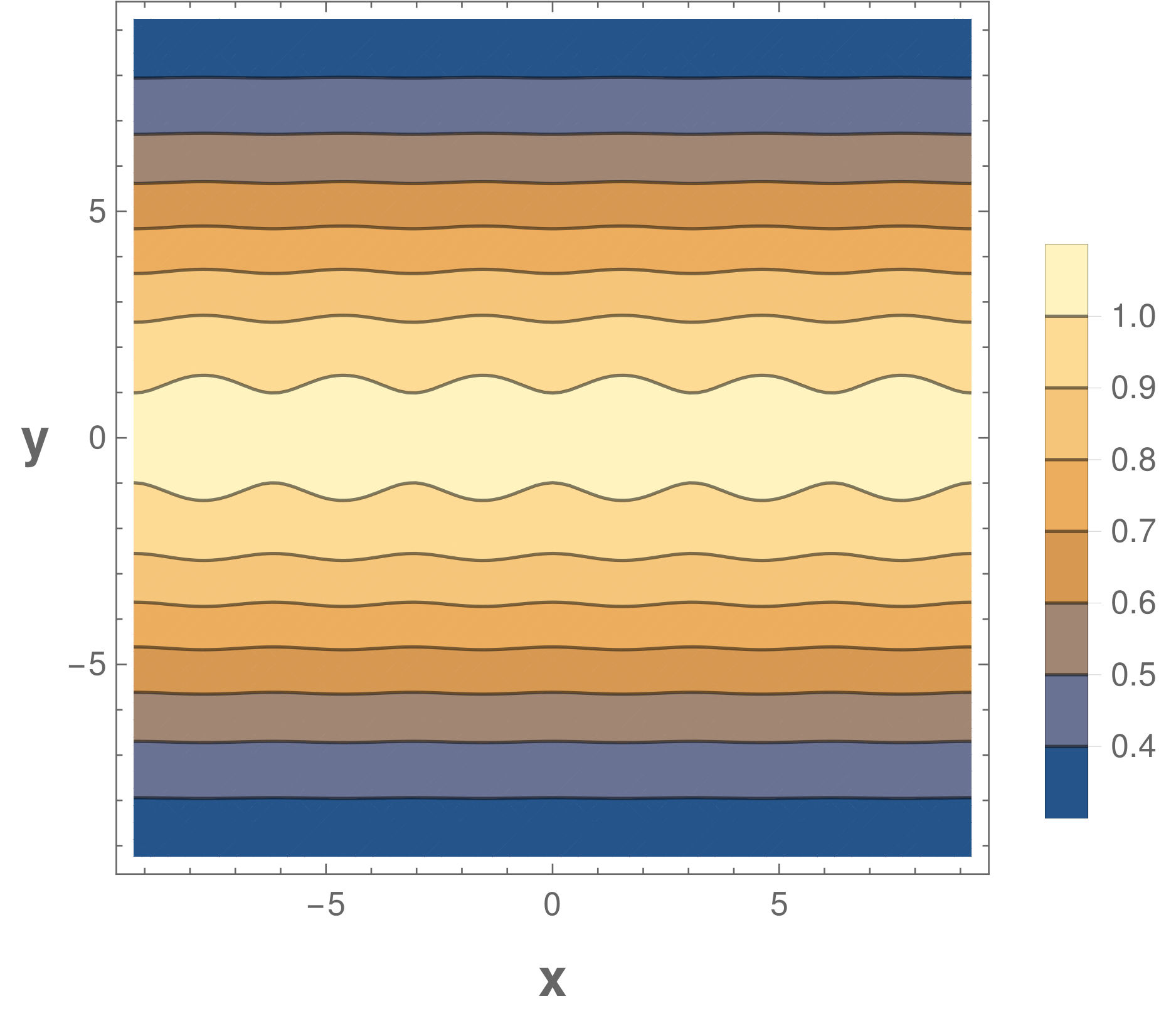}	
		\includegraphics[scale=0.32]{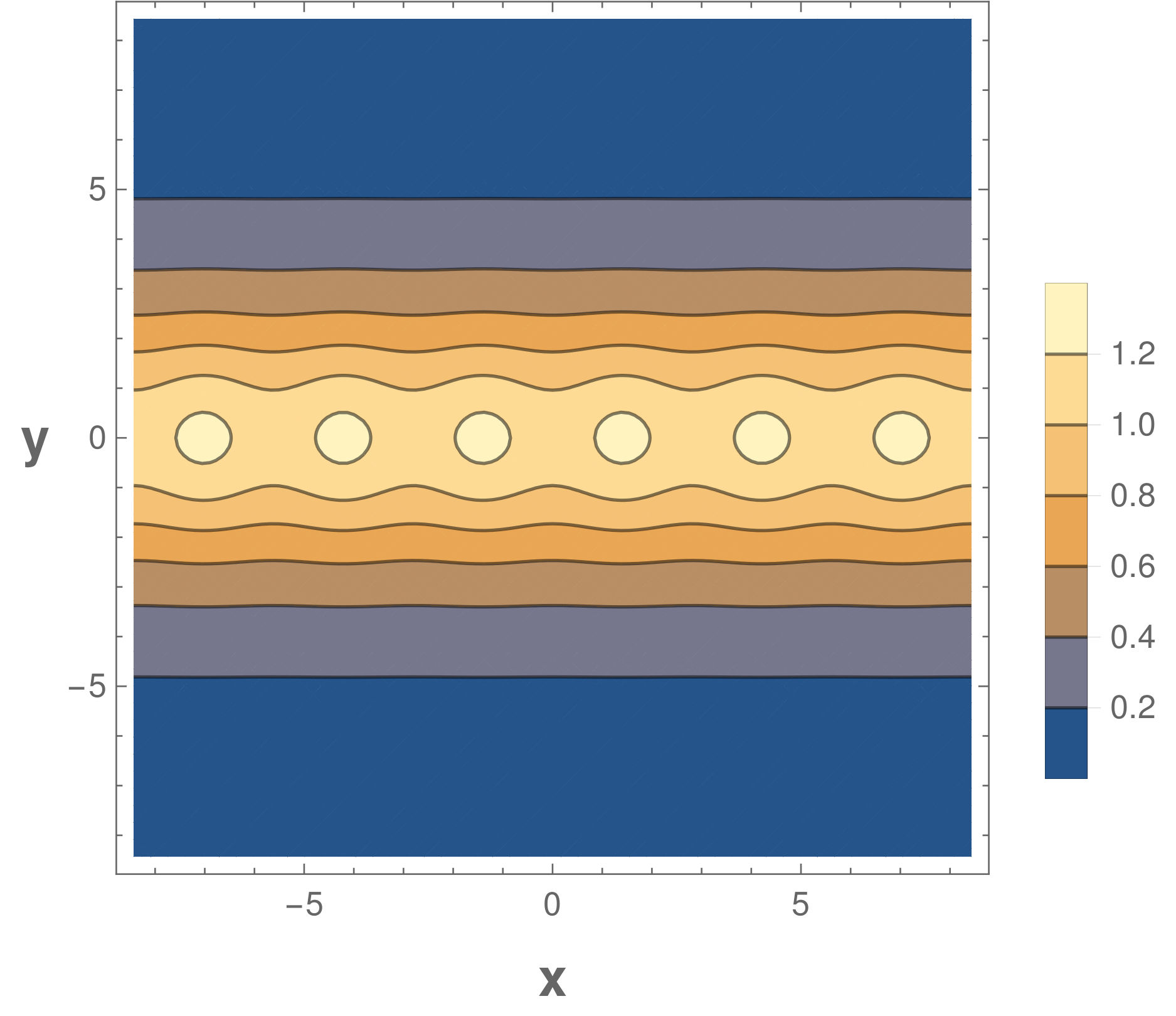}	\includegraphics[scale=0.32]{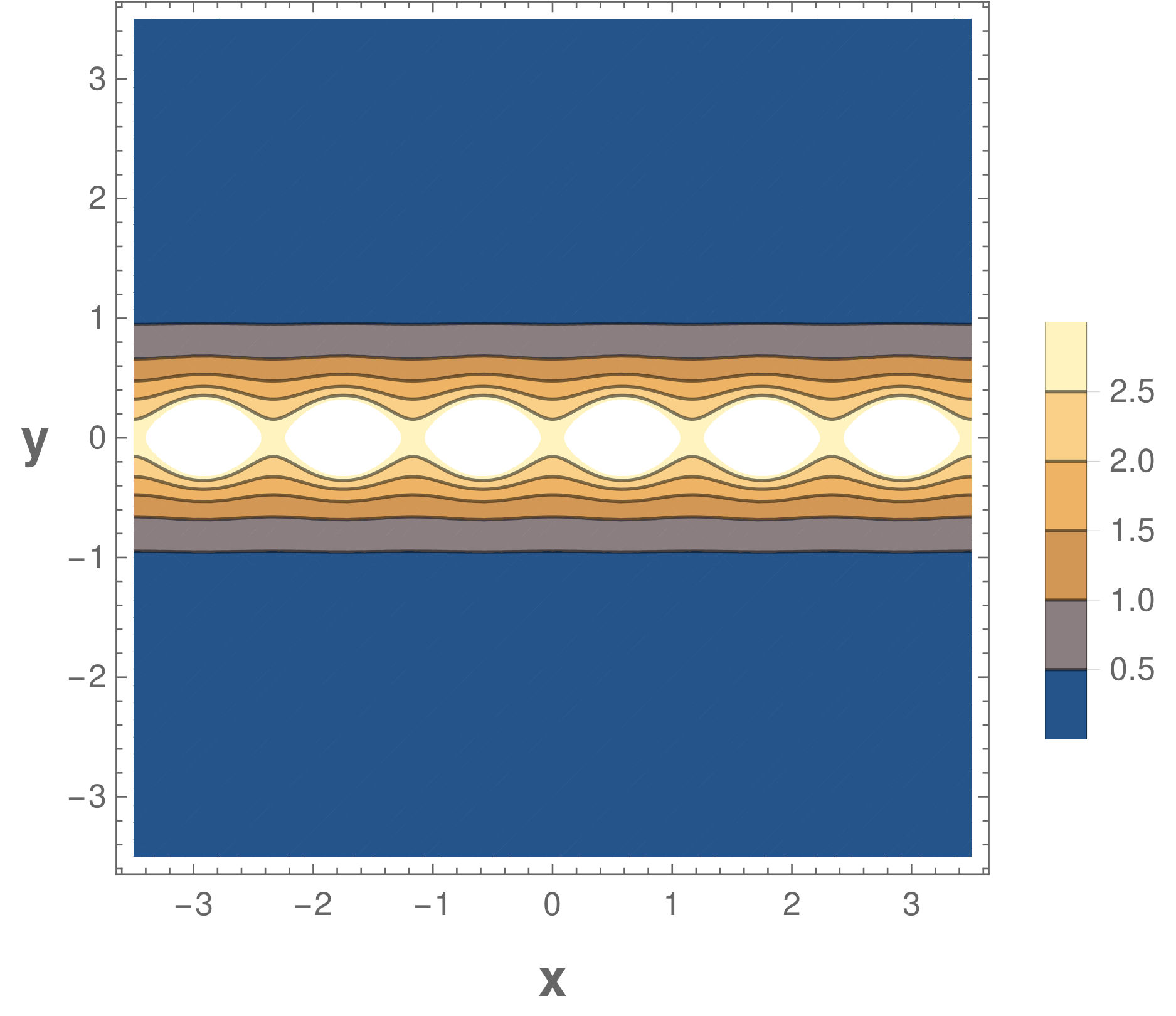}\\
		$\omega=0.2$ \hspace{40mm} $\omega=0.5$\hspace{40mm} $\omega=2.5$
	\end{center}
	\caption{(color online) The potential term $\mathcal{V}_A(x,y)$ (upper row) with density of probability of the confined state $\widetilde{v_1}$  (lower row). The columns differ by the choice of $\omega$. In all plots, we used $m=1$, and $\phi_1=\frac{\pi}{15}$, $\phi_2=\frac{\pi}{2.1}$. Densities of probability of $\widetilde{v}_1$ and $\widetilde{v}_2$ are virtually indistinguishable. The functions went over the cutoff in the purely white dots in the right column.
	}\label{modelA}
\end{figure}

The Wick rotation transforms the free-particle Dirac operator $H_0$ into the two-dimensional Hamiltonian of a massless particle,
\begin{equation}\label{WRH0}
\widetilde{H}_0=-i\sigma_1\partial_x-i\sigma_2\partial_y+m\sigma_0.
\end{equation}
The mass term of (\ref{h1}) is converted into the electrostatic potential. Following (\ref{2d2a}), we get this stationary equation,
\begin{equation}\label{V1}
\widetilde{H}_1\widetilde\psi(x,y)=\left(-i\sigma_2\partial_y-i\sigma_1\partial_x+\widetilde{V}(x,y)\sigma_0\right)\widetilde{\psi}(x,y)=0,
\end{equation}
where the potential term is
\begin{equation}
\widetilde{V}(x,y)=m-\frac{4mk^2\cosh^2 (\omega y)}{m^2+k^2\cosh (2\omega y)+\omega^2\cos (2k x)}.
\end{equation}
For large $y$, it rapidly converges to a constant value,
$\lim_{y\rightarrow\pm\infty}\widetilde{V}(x,y)=-m.$
Let us substract this asymptotic value from the electrostatic potential and attribute it to the energy of the fermion. This way, we get the new potential $\mathcal{V}_A$, 
\begin{equation}\label{Vm}
\mathcal{V}_A(x,y)=\widetilde{V}(x,y)+m=- \frac{4m\omega^2\sin^2(k x)}{m^2+k^2\cosh(2\omega y)+\omega^2\cos(2kx)}.
\end{equation}
It vanishes asymptotically for large $|y|$. Its sign depends on the sign of $m$ and it is periodic in $x$.
 It is also even with respect to reflection,
\begin{equation}
\lim_{|y|\rightarrow\infty}\mathcal{V}_A(x,y)=0,\quad \mathcal{V}_A(x+T,y)=\mathcal{V}_A(x,y), \quad \mathcal{V}_A(x,y)=\mathcal{V}_A(-x,-y),
\end{equation}
where $T=\pi/k$. It forms a thin electrostatic grating or a comb, see Fig.~\ref{modelA}. 
Substituting (\ref{Vm}) into (\ref{V1}), we get the stationary equation 
for energy $E=m$, 
\begin{equation}\label{V1b}
\left(-i\sigma_1\partial_x-i\sigma_2\partial_y+\mathcal{V}_A(x,y)\sigma_0\right)\widetilde\psi(x,y)=m\widetilde{\psi}(x,y).
\end{equation}

The localized solutions (\ref{G}) get transformed into $\widetilde{v}_1$ and $\widetilde{v}_2$ via (\ref{tildepsi}),
\begin{equation}
\widetilde{v}_1=\frac{\sqrt{2}k^2}{\widetilde{D}_1}\left(\begin{array}{c}-i\cosh(\omega y)[k\cos (kx)+m\sin (kx)]+\omega \sin (kx)\sinh (\omega y)\\\cosh (\omega y)[-k\cos (k x)+m\sin (kx)]+i\omega \sin( kx)\sinh (\omega y)\end{array}\right),
\end{equation}
\begin{equation}
\widetilde{v}_2=\frac{1}{\sqrt{2}\widetilde{D}_1}\left(\begin{array}{c}\cosh (\omega y)[k\cos (k x)-m\sin (kx)]+i\omega \sin (kx)\sinh (\omega y)\\i\cosh(\omega y)[k\cos (kx)+m\sin (kx)]+\omega \sin (kx)\sinh (\omega y)\end{array}\right).
\end{equation}
We abbreviated here $\widetilde{D}_1=m^2+\omega^2\cos (2kx)+k^2\cosh (2\omega y)$. Brief inspection of these formulas reveals that these states are exponentially vanishing for large $|y|$. Therefore, these solutions correspond to the states strongly localized by the electrostatic potential, see Fig \ref{modelA}.

\begin{figure}[h!]
	\begin{center}
		\includegraphics[scale=.32]{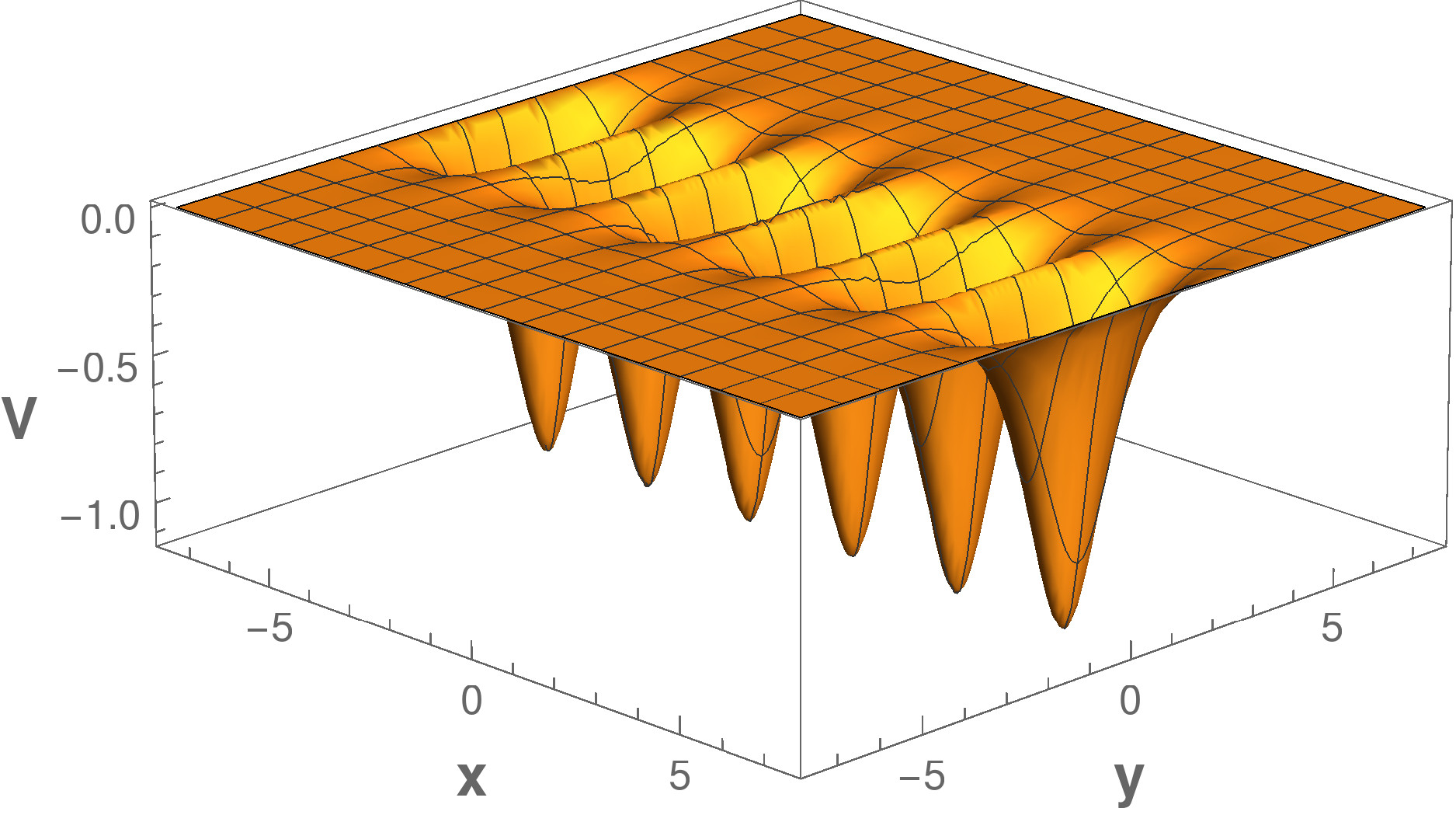}
		\includegraphics[scale=.32]{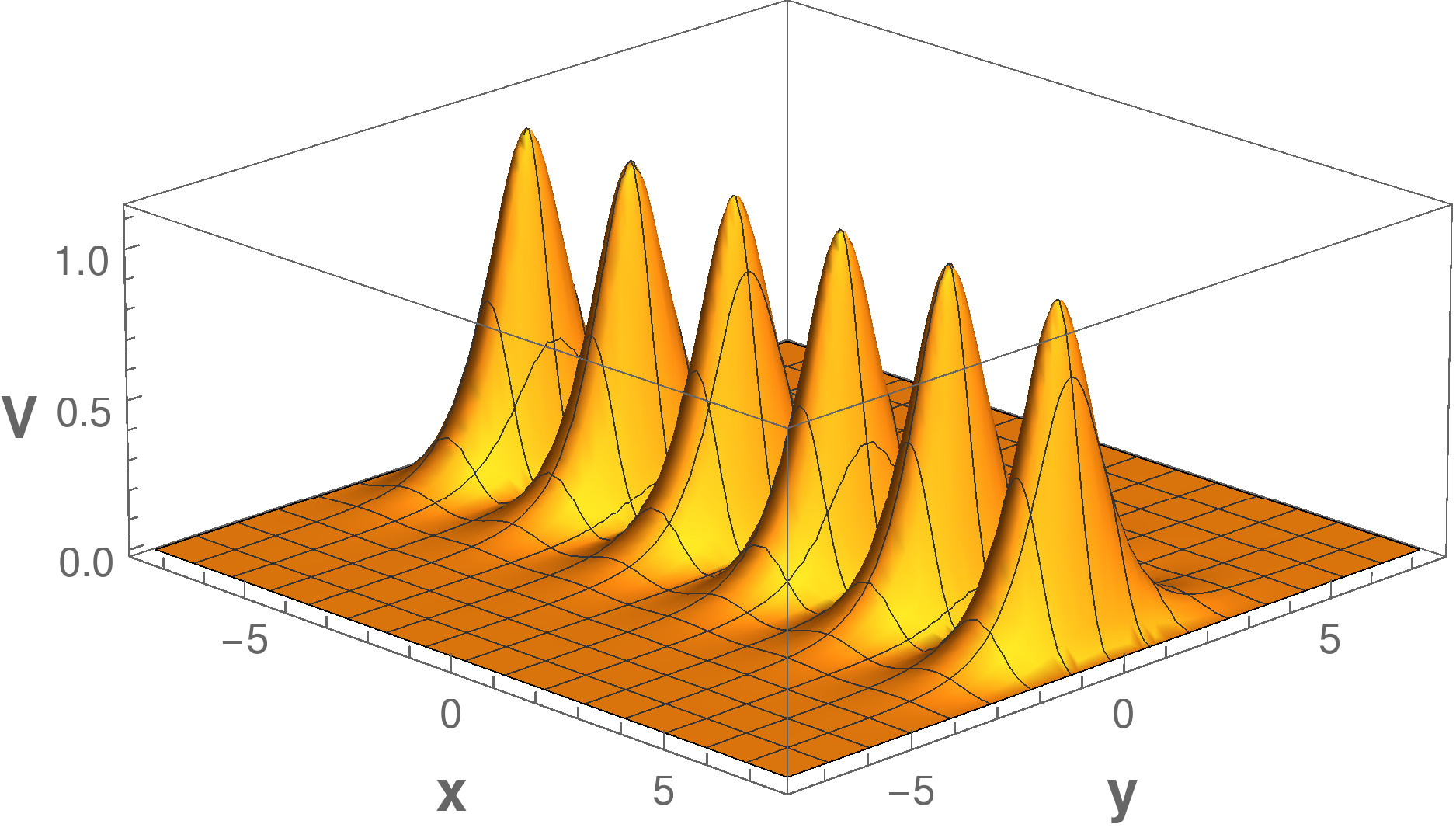}\\
		$ $ \hspace{38mm}  $ $\\
		\includegraphics[scale=.32]{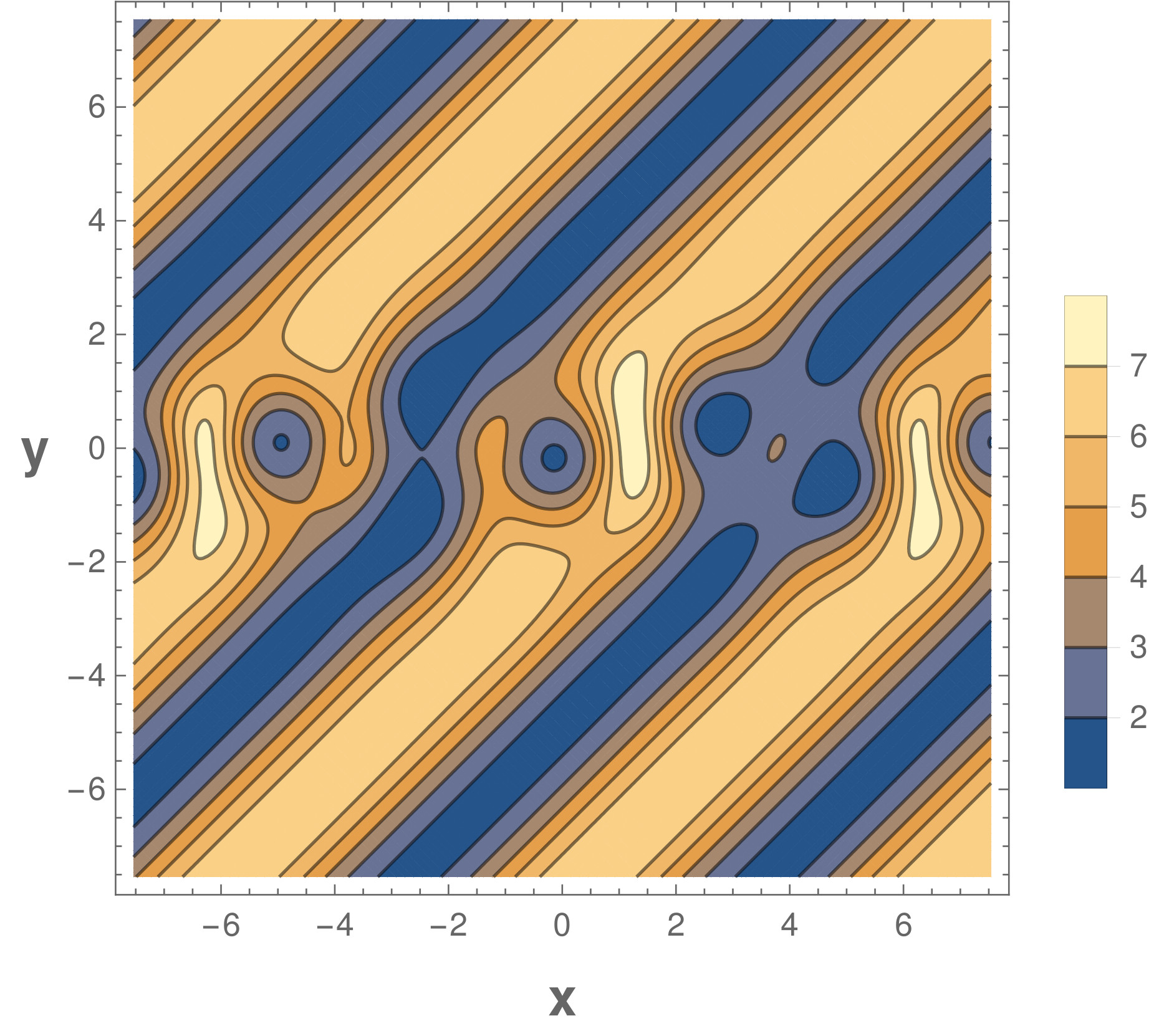}
		\includegraphics[scale=.32]{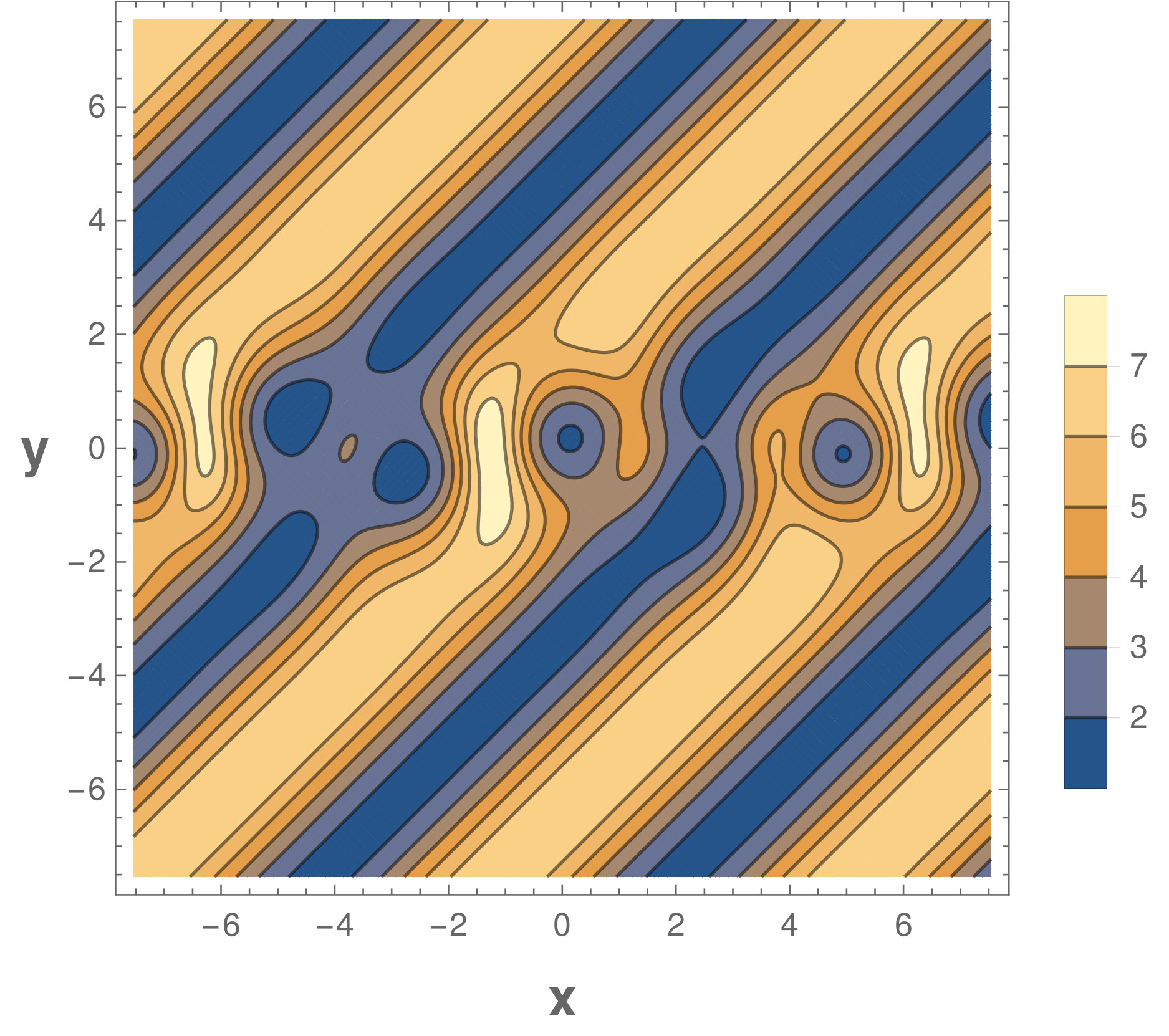}\\
		$m=1$ \hspace{42mm}  $m=-1$
	\end{center}
	\caption{(color online) The potential term $\mathcal{V}_A(x,y)$ is plotted in the upper row. In the lower row, we show the density of probability (interference pattern) of a linear combination of a free particle (left) and the asymptotically plane-wave solutions (\ref{Superposition2}) for the upper row potential for  $m=1$ (left) and $m=-1$ (right). In all plots, we used $\omega=0.75$, and $\phi_1=0$, $\phi_2=\frac{\pi}{2}$. 
	}\label{modelAb}
\end{figure}

The electrostatic potential $\mathcal{V}_A(x,y)$ is periodic in $x$, but it ceases to have translation invariance. Any plane wave incoming from the left will hit the barrier under non-normal direction locally and the back-scattering can be expected. However, the potential is reflectionless for the states with energy $E=m$. They are just phase-shifted when passing through the potential barrier. We can show it with the use of the interwtining operator $\widetilde{L}$ that maps the free-particle solutions of (\ref{WRH0}) into the solutions of (\ref{V1}), see (\ref{tildeL}). The operator $\widetilde{L}$ acquires the following explicit form 
\begin{eqnarray}\label{tildeL1}
\widetilde{L}&=&e^{i\frac{\pi}{4}\sigma_1}L(ix,y)e^{-i\frac{\pi}{4}\sigma_1}\nonumber \\
&=&-i\partial_x-\frac{2k^2}{\widetilde{D}_1}\left(\frac{i\,\omega^2\sin (2kx)}{2k}\sigma_0-m\cosh^2(\omega y)\, \sigma_1-\frac{\omega\, \sinh (2\omega y)}{2}\sigma_3\right). 
\end{eqnarray} We fix the wave vector $\vec{k}$ for an incoming plane wave as follows
\begin{equation}
k_x=m\sin\phi,\quad k_y=m\cos\phi,\quad\phi=(-\pi/2,\pi/2).
\end{equation}
Then the free-particle solutions of $\widetilde{H}_0\widetilde{\psi}_0(x,y,\phi)=0$ can be written as
\begin{eqnarray}\label{scatt0}
&&\widetilde{\psi}_0(x,y,\phi)=e^{i m(\sin\phi\, x+\cos\phi\,y)}\left(\begin{array}{c}1\\-i e^{-i\phi}\end{array} \right).
\end{eqnarray}
The parameter $\phi$ corresponds to the incidence angle of the plane wave. 
The solution $\widetilde{\psi}_0$ can be transformed into the scattering state of (\ref{V1b}),
\begin{equation}
\left(-i\sigma_1\partial_x-i\sigma_2\partial_y+\mathcal{V}_A(x,y)\right)\widetilde{L}\widetilde\psi_0(x,y)=m\widetilde L\widetilde{\psi}_0(x,y).
\end{equation}
Let us see how the scattering state $\widetilde{L}\widetilde{\psi}_0$ is affected when passing through the potential barrier. 
For large $|y|$, the operator $\widetilde{L}$ has the following asymptotic form
\begin{equation} \label{asymtotic}
\widetilde{L}\sim\begin{cases}-i\partial_{x}+m\sigma_1-\omega\sigma_3,\quad y\rightarrow-\infty,\\-i\partial_{x}+m\sigma_1+\omega\sigma_3,\quad y\rightarrow \infty.\end{cases}
\end{equation} 
Hence, the action of $\widetilde{L}$ on $\widetilde{\psi}_0$ is 
\begin{equation}\label{superklein}
\lim_{y\rightarrow\pm\infty}\widetilde{L}\widetilde{\psi}_0=e^{i m(\sin\phi\, x+\cos\phi\,y)}(\pm \omega-im\cos\phi)\left(\begin{array}{c}1\\i e^{-i\phi}\end{array}\right).
\end{equation}
The relation (\ref{superklein}) reveals that there is no backscattering on the potential, independently on the incidence angle $\phi=\arctan\frac{k_{x}}{k_{y}}$. Hence, there emerges the omnidirectional tunneling of Dirac fermions through the barrier for $E=m$. The scattering state $\widetilde{L}\widetilde{\psi}_0$ accumulates the phase shift when passing through the barrier,
\begin{equation}\label{phaseshift1}
\lim_{y\rightarrow+\infty}\widetilde{L}\widetilde{\psi}_0=\frac{ \omega-im\cos\phi}{ -\omega-im\cos\phi}\left(\lim_{y\rightarrow-\infty}\widetilde{L}\widetilde{\psi}_0\right).
\end{equation}
The phase shift depends both on the potential (determined by the values of $m$ and $\omega$) and on the incidence angle $\phi$. It is symmetric with respect to $\phi\rightarrow -\phi$ which is to be expected as the potential term is symmetric with respect to $x\rightarrow-x$.  It changes  the interference pattern of a linear combination of the plane waves. Taking the sum of two plane waves with incidence angles $\phi_1$ and $\phi_2$, 
\begin{equation}\label{Superposition}
F_A(x,y,\phi_1,\phi_2)=\widetilde{L}\widetilde{\psi}_0(x,y,\phi_1)+\widetilde{L}\widetilde{\psi}_0(x,y,\phi_2),
\end{equation} 
the interference pattern gets slightly shifted along $x$ axis when passing through the potential barrier, see Fig.~\ref{modelAb} for illustration.

\subsection{Wick-rotated model B}
Let us turn our attention to the model B described by (\ref{h2}). Likewise in the model A, the Wick rotation transforms the mass term of $H_1$ into the electrostatic potential $\widetilde{V}_1=\left(m + W(i x, y)  \right)$ of the Wick-rotated operator $\widetilde{H}_1$,
\begin{equation}
\widetilde{H}_1=-i\sigma_1\partial_x-i\sigma_2\partial_y+\widetilde{V}_1(x,y)\sigma_0.
\end{equation} 
Revising the explicit form of $W$ in (\ref{W2}), we can see that the substitution $z\rightarrow ix$ makes it complex-valued and the operator $\widetilde{H}_1$ results to be non-Hermitian in general.  Fortunately, we can fix this issue by setting the parameter $s^2= \omega^4/k^4+4(c_1^2+c_2^2)k^2$. Then the potential term becomes
\begin{eqnarray}
\widetilde{V}_1= \left( m + 
\frac{ k^2\left[ 2ms +4 c_1 k^2 - 2 m(2 c_1 m + s) \cosh(2\omega y) + 4 c_2 m \omega \sinh(2 \omega y)  \right]}{\widetilde{D}_2} \right) ,
\end{eqnarray}
where
\begin{eqnarray}\label{D}
\widetilde{D}_2(x,y)= \omega^4\left(\cos(2 k x)- \frac{k^2}{\omega^4}m(m s + 2 c_1 k^2)\right) +k^4  \left[ (2 c_1 m + s) \cosh(2 \omega y)-2 c_2 \omega \sinh(2\omega y)   \right].
\end{eqnarray}
Besides Hermiticity, we should also guarantee that the new potential $\widetilde{V}_1$ is regular. Let us discuss the range of the parameters  $m$, $\omega$, $c_1$ and $c_2$ where this requirement can be met. We shall find the lower bound for $\widetilde{D}_2$ and require it to be positive. The second term in (\ref{D}) is a function of $y$,
\begin{equation}
f(y)=k^4((2 c_1 m + s) \cosh(\omega y)-2 c_2 \omega \sinh(2\omega y)),
\end{equation}
and it has its minimum $f_{min}=k^4\sqrt{(2c_1m+s-2c_2\omega)(2c_1m+s+2c_2\omega)}$ at $y=\frac{1}{2\omega}\log\sqrt{\frac{2c_1m+s+2c_2\omega}{2c_1m+s-2c_2\omega}}$. The minimum $f_{min}$ has to be real-valued. We can guarantee it by demanding
\begin{equation}\label{ineq1}
2c_1m+s> |2c_2 \omega|.
\end{equation}
We can find the lower bound $\widetilde{D}_{min}$ for $\widetilde{D}_2$ by replacing $f(y)$ and $\cos(2kx)$ by their minimal values. We require it to be strictly positive, 
\begin{equation}\label{ineq2}
\widetilde{D}_2\geq \widetilde{D}_{min}=-\omega^4\left(1+ \frac{k^2}{\omega^4}m(m s + 2 c_1 k^2)\right)+ f_{min}>0.
\end{equation}
As $\widetilde{D}_{min}$ depends on the eligible real parameters $m$, $\omega$, $c_1$ and $c_2$ only, the inequalities (\ref{ineq1}) and (\ref{ineq2}) form sufficient conditions for regularity of the potential term $\widetilde{V}_1$. In the numerical tests that we performed, it was rather problematic to find a setting that would {\it not} comply with (\ref{ineq1}) and (\ref{ineq2}), aside from $c_1=c_2=0$ where $D_{min}=0$. 

\begin{figure}\begin{center}
		\includegraphics[scale=.31]{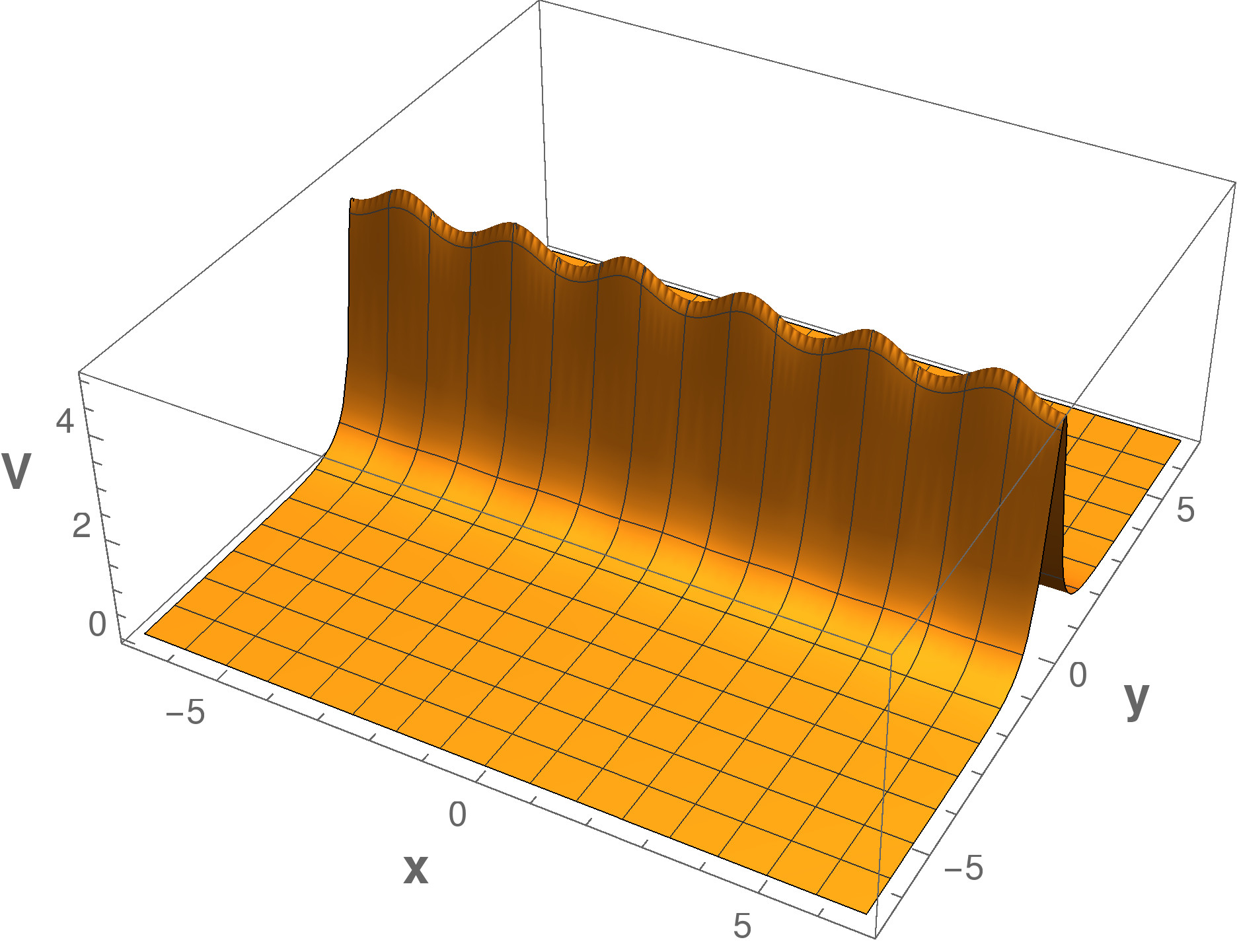}
		\includegraphics[scale=0.31]{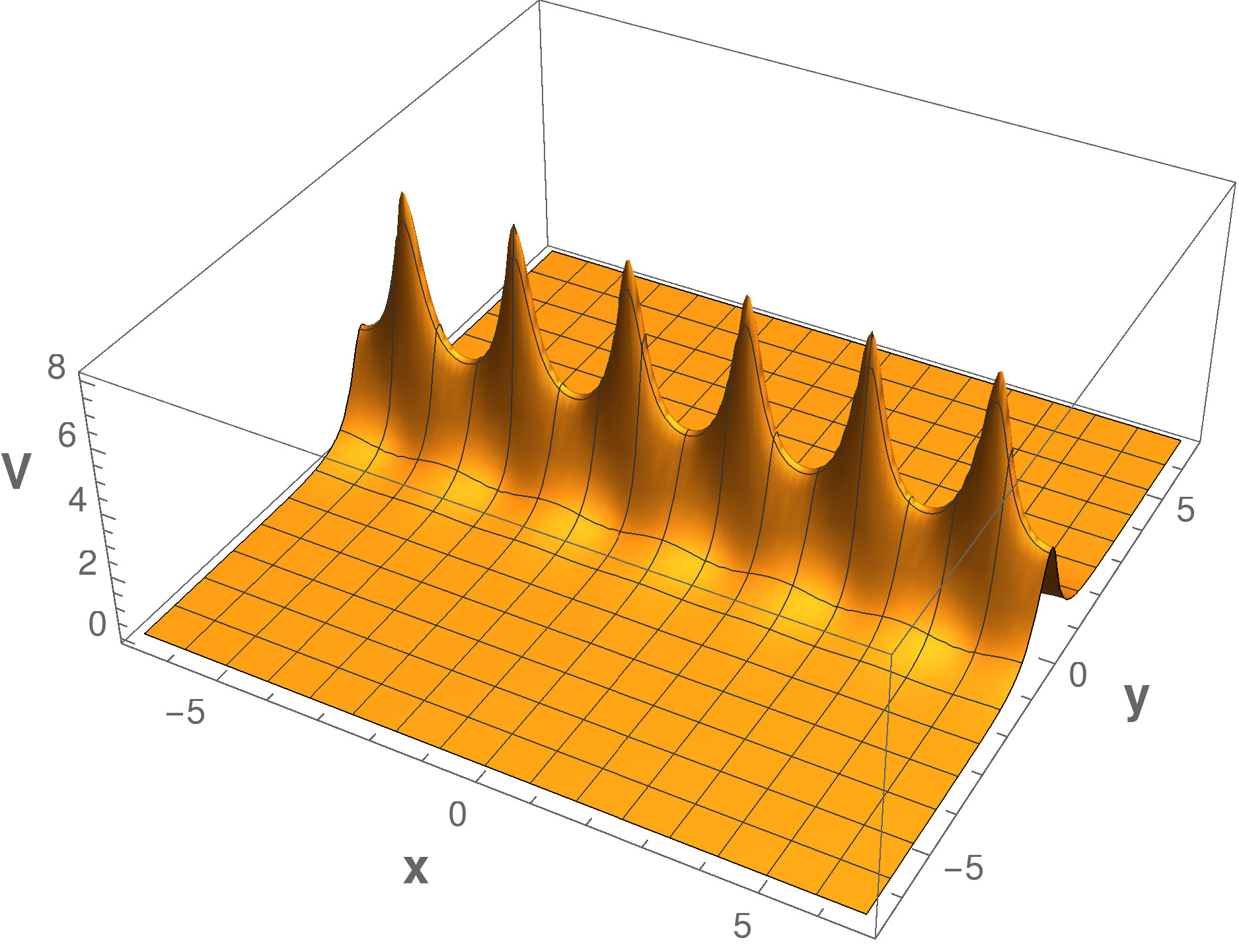}
		\includegraphics[scale=.31]{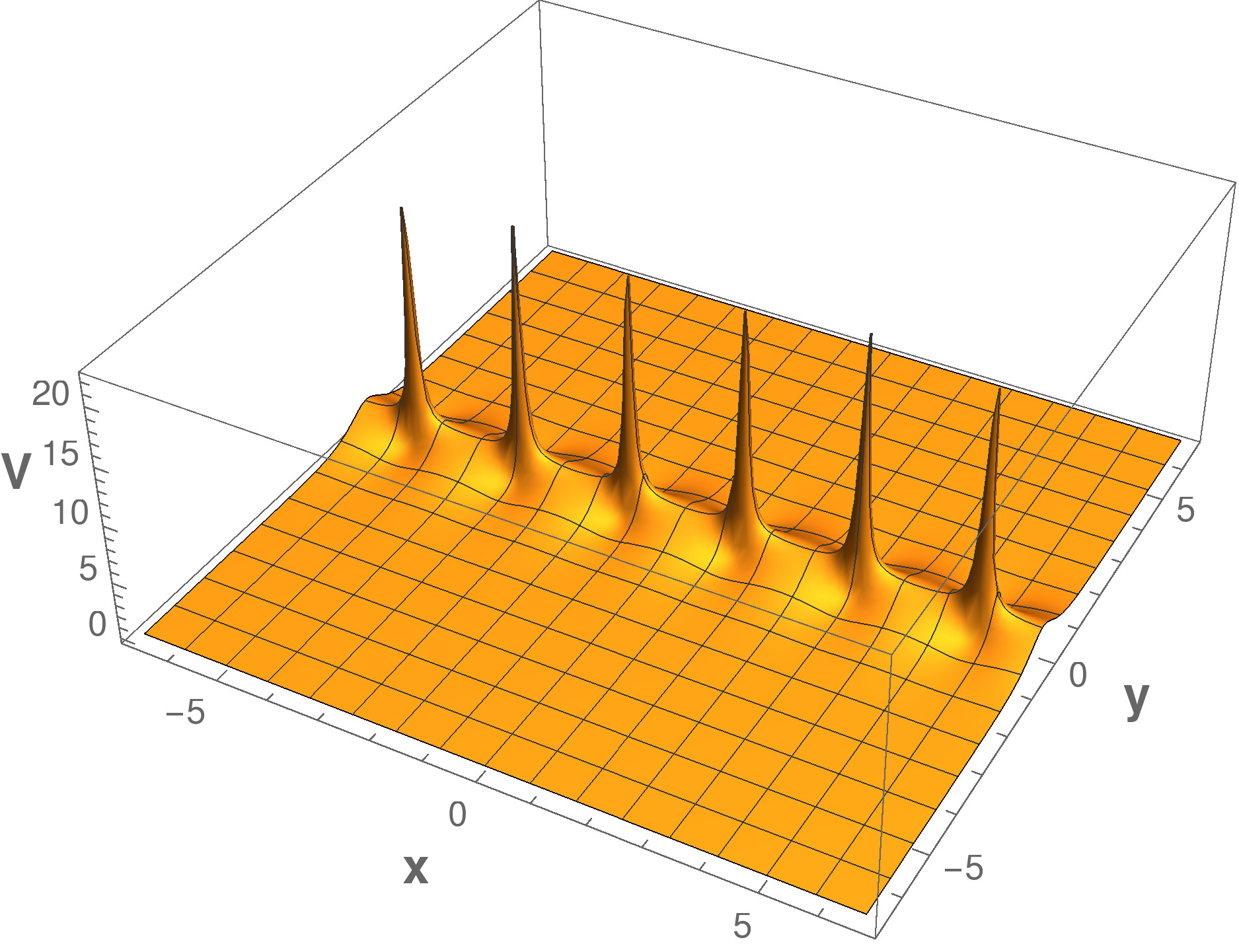}\\
		$ $ $ $\\	\includegraphics[scale=0.31]{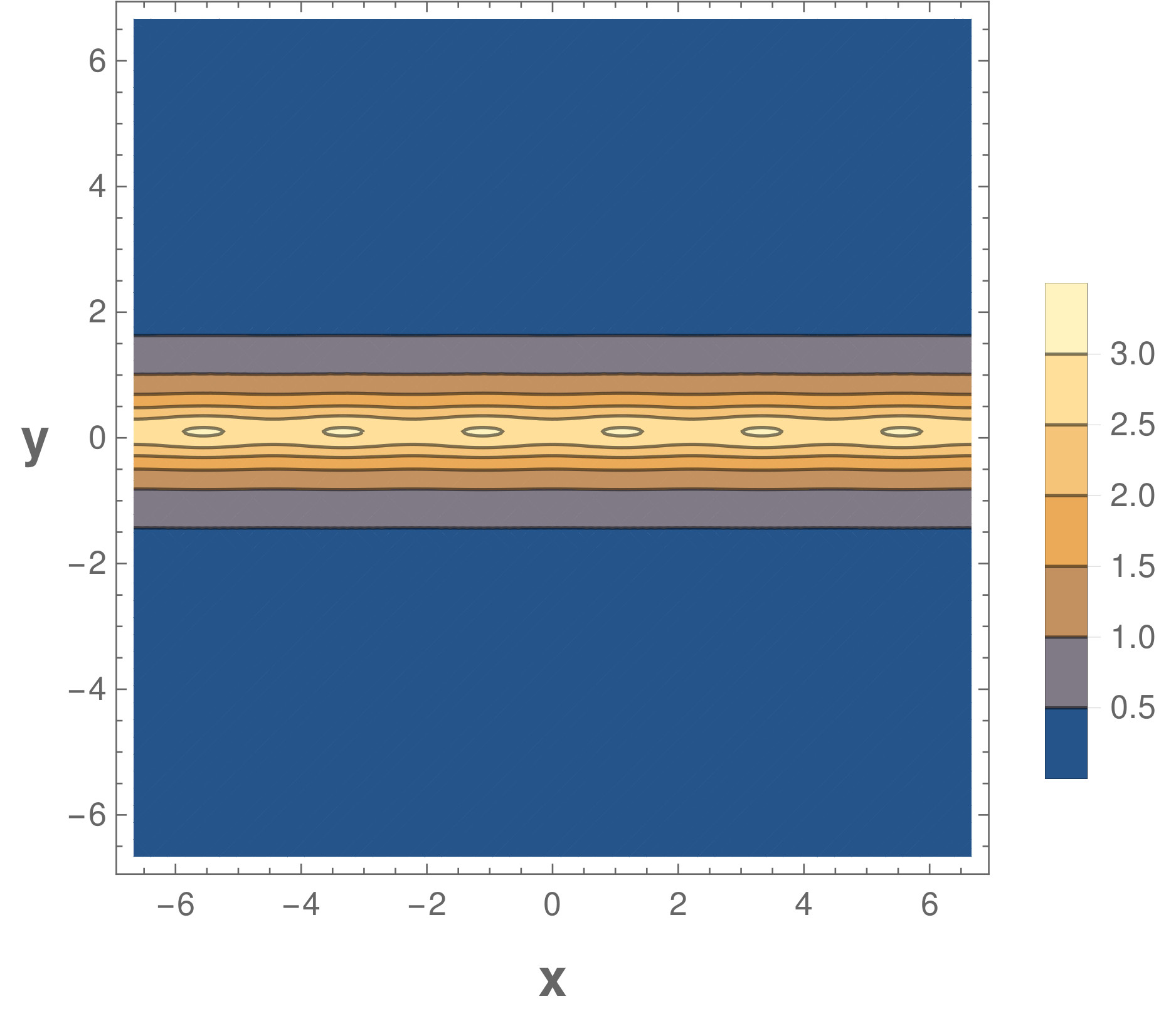}	
		\includegraphics[scale=0.31]{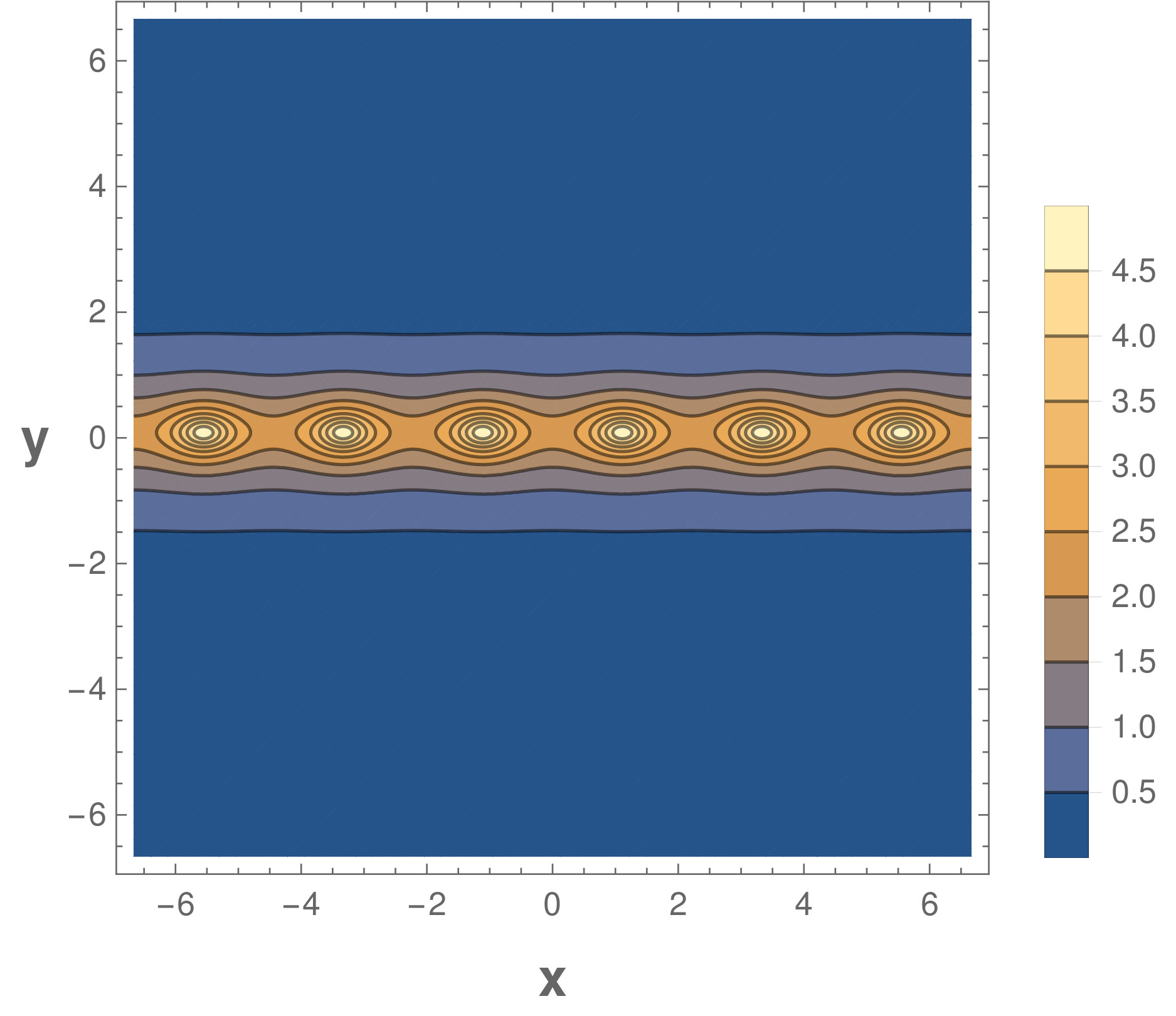}	\includegraphics[scale=0.31]{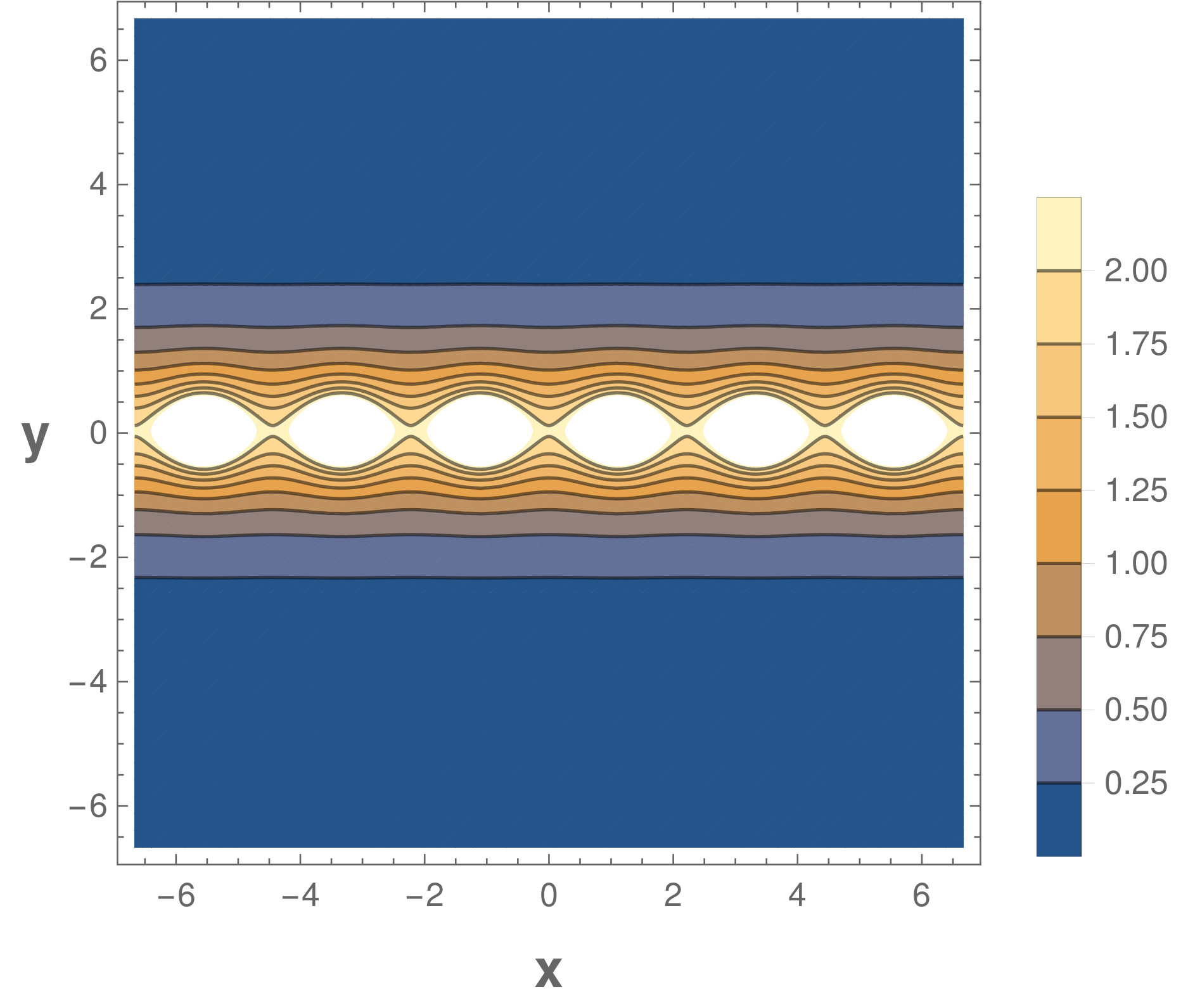}\\
		$ $ $ $\\
		\includegraphics[scale=0.31]{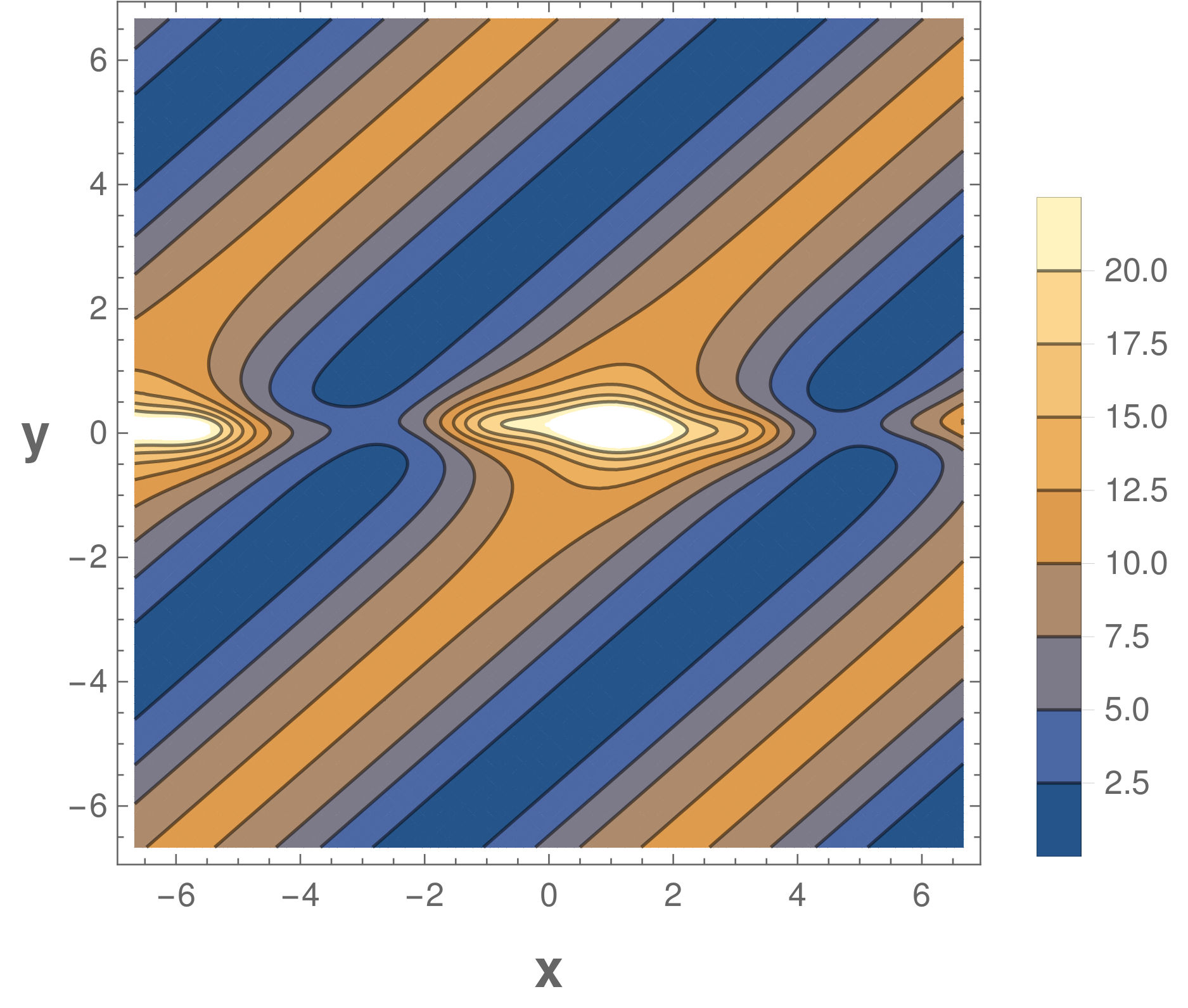}	
		\includegraphics[scale=0.31]{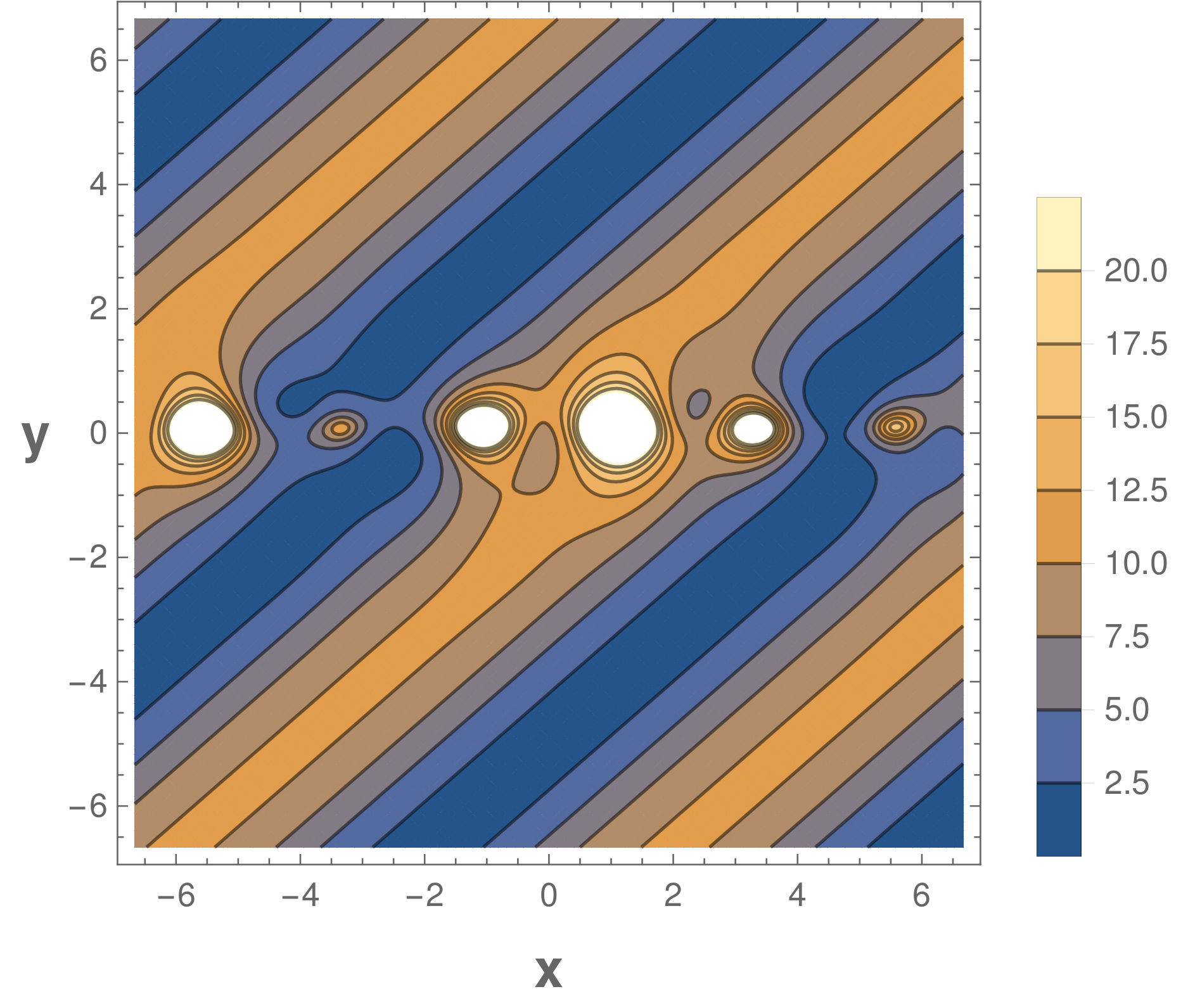}	\includegraphics[scale=0.31]{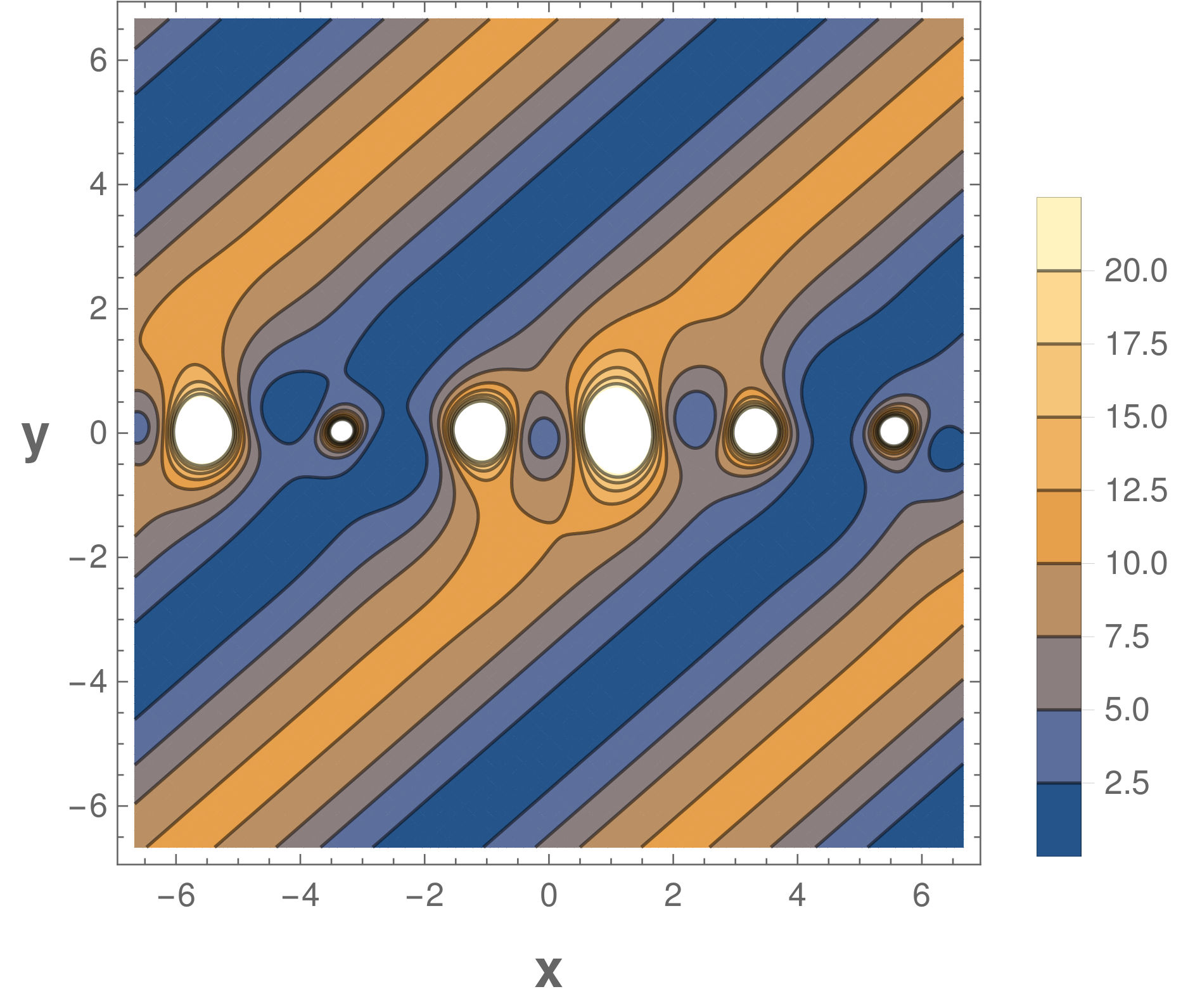}\\
		$c_1=2,\ c_2=1$ \hspace{25mm} $c_1=0.2,\ c_2=0.1$\hspace{25mm} $c_1=0.04,\ c_2=0.02$
	\end{center}
	\caption{(color online) The potential term $\mathcal{V}_B(x,y)$ (upper row) with density of probability of the confined states $\widetilde{v_1}$  (central row) and the density of probability (interference pattern) of a linear combination of the asymptotically plane-wave solutions (\ref{Superposition2}) (lower row). The columns differ by the choice of $c_1$ and $c_2$. In all plots, we used $m=1$, $\omega=1$ and $\phi_1=\frac{\pi}{15}$, $\phi_2=\frac{\pi}{2.1}$. Densities of probability of $\widetilde{v}_1$ and $\widetilde{v}_2$ are virtually indistinguishable. The functions went over the cutoff in the pure white dots in the lower row.
	}\label{modelB}
\end{figure}

The potential $\widetilde{V}_1$ tends exponentially towards $-m$  for large $|y|$. Likewise in the previous case, we subtract the asymptotic value and define the new function $\mathcal{V}_B$,
\begin{equation}\label{Vm2}
\mathcal{V}_B=\widetilde{V}_1+m.
\end{equation}
The new potential $\mathcal{V}_B$ satisfies
\begin{equation}
\lim_{|y|\rightarrow\infty}\mathcal{V}_B(x,y)=0, \quad \mathcal{V}_B(x+T,y)=\mathcal{V}_B(x,y)
\end{equation}
for  $T=\frac{\pi}{k}$.  This way, (\ref{V1}) turns into the stationary equation for a massless particle of energy $E=m$,
\begin{equation}\label{V2b}
\left(\-i\sigma_1\partial_x-i\sigma_2\partial_y+\mathcal{V}_B(x,y)\sigma_0\right)\widetilde\psi(x,y)=m\widetilde{\psi}(x,y).
\end{equation}
The potential $\mathcal{V}_B$ forms a periodic chain of scatterers that are interconnected by a barrier. The periodicity of the chain can be controlled by $\omega$ and $m$ (recall $k=\sqrt{\omega^2+m^2}$).  Having these parameters fixed, the height of the intermediate barrier relative to the peaks of the scatterers can be controlled by $c_1$ and $c_2$, see Fig.~\ref{modelB}. The potential can be both attractive and repulsive in dependence on the choice of the parameters.

The solutions $\widetilde{v}_1$ and $\widetilde{v}_2$ can be obtained directly from (\ref{G2}) by the transformation (\ref{tildepsi}). Their explicit form is not quite compact so that we prefer not to present them here explicitly. Instead, we refer to the Fig.~\ref{modelB} where the corresponding density of probability is illustrated. We can see that these states are strongly localized by the potential.

Similarly to the previous model, there also occurs super-Klein tunneling at $E=m$. We can show it with the use of the intertwining operator $\widetilde{L}$ that maps free-particle solutions into the solutions of (\ref{V2b}).  It can be obtained from (\ref{LmodelB}) as
\begin{equation}\label{LmodelBwicked}
\widetilde{L} = -i\partial_x - \frac{1}{D_2(ix,y)}\left( \ell_0(ix)\sigma_0 + \ell_1(y) \sigma_1 - \ell_2(y) \sigma_3 \right)
\end{equation}

For large $|y|$, the intertwining operator (\ref{LmodelBwicked}) has the same asymptotic form \eqref{asymtotic} as the intertwining operator of in the model A. Therefore, there is no backscattering on the potential barrier and
we get exactly the same phase shift as in (\ref{phaseshift1}), despite the fact that the electrostatic potentials are rather different.
Let us illustrate the effect of the phase shift on the superposition 
\begin{equation}\label{Superposition2}
F_B(x,y,\phi_1,\phi_2)=\widetilde{L}\widetilde{\psi}_0(x,y,\phi_1)+\widetilde{L}\widetilde{\psi}_0(x,y,\phi_2).
\end{equation} 
Likewise in the previous model, the electrostatic potential shifts the interference pattern along the $x$-axis, see Fig.~\ref{modelB}.

\section{Discussion}

To our best knowledge, the super-Klein tunneling was discussed up to now for electric fields that possessed translation symmetry and, therefore, the corresponding stationary equations were separable. We presented here two models where the electrostatic barrier was truly two-dimensional. The scattering of Dirac fermions on such potentials can be profoundly more complicated when compared to the models with translation symmetry. The incident wave bounces on the barrier with angle that varies locally so that non-trivial interference of the incoming and reflected wave is to be expected. This is, without any doubt, also the case in our models. However, for $E=m$, the interference is such that the reflected wave gets completely annihilated and the waves incoming  from any direction get perfectly transmitted up to a phase shift.

The current work follows the spirit of \cite{disguise}. There, the total transmission of massless fermions bouncing on the electrostatic barrier for a fixed angle (normal incidence) was explained via unitary transformation that related the model with the free particle scattering. The solutions corresponding to the fermions incoming in normal direction were mapped to the free-particle solutions and, therefore, no back-scattering could occur. In the current work, the stationary equation for a specific fixed energy was related to the free-particle one by the intertwining operator (\ref{tildeL}). This link allowed us to show that at that specific energy, there is no back-scattering on the potential independently on the incidence angle so that super-Klein tunneling occurs. We also showed that there also exist states that are strongly localized by the electrostatic barrier.

The electrostatic potentials $\mathcal{V}_A$ in (\ref{Vm}) and $\mathcal{V}_B$ in (\ref{Vm2}) of the Wick-rotated models A and B have quite different form, compare Fig.~\ref{modelA} with Fig.~\ref{modelB}. With $m$ and $\omega$ fixed, the shape of the potential $\mathcal{V}_B$ can be tuned additionally by the parameters $c_1$ and $c_2$. However, the formulas for the phase shifts in the Wick-rotated model A and model B \textit{are identical}.  It means that whatever are the potentials for given $m$ and $\omega$, they are indistinguishable from their long-range effect on the scattered waves.  We cannot see any simple explanation for this remarkable fact. Nevertheless, a possible guidance could be provided by the results published in \cite{AKNS,dunnethies, twisted} on the reflectionless integrable models described by the one-dimensional Dirac equation. Having in mind connection of the later systems with the integrable fermion models in $1+1$ dimensions \cite{dunnethies,twisted}, it would be interesting to see whether the settings described here could play such a role in higher dimensions. However, we find deeper analysis of these questions out of the scope of the current article.

\section*{Acknowledgements}
FC was partially supported by Fondecyt grant 1171475 and Becas Santander Iberoam\'erica. FC would like to thank the Departamento de F\'isica Te\'orica, \'Atomica y \'Optica at the Universidad de Valladolid and the Department of Theoretical Physics of the Nuclear Physics Institute of the Czech Academy of Science in \v Re\v z for all the support and kind hospitality.
VJ was supported by GA\v CR grant no.19-07117S. The authors would like to thank to the 
Centro de Ciencias de Benasque Pedro Pascual where this project was initiated.

\end{document}